\documentclass[letter]{aa}  
\pdfoutput=1
\bibpunct{(}{)}{;}{a}{}{,} 
\usepackage[utf8]{inputenc}
\usepackage[T1]{fontenc}
\usepackage{textcomp}
\usepackage{graphicx}
\usepackage{multirow}
\usepackage{pifont}
\usepackage{makecell} 
\setcellgapes{5pt}
\usepackage{txfonts}
\usepackage{rotating}
\usepackage[]{xcolor}
\usepackage{comment}
\usepackage{amsmath,bm}
\usepackage{booktabs}
\usepackage{nicematrix}
\usepackage{hyperref}

\newcommand{\hii}{\mbox{H\,{\sc{ii}}\,}}
\usepackage{graphicx}
\usepackage{txfonts}
\usepackage{tikz}
\usepackage{ulem}
\usepackage{subfigure}
\usepackage{ifthen}
\usepackage{placeins}
\usepackage{lipsum}
\usepackage{float}

\hypersetup{colorlinks=true,linkcolor=blue,citecolor=blue,filecolor=blue,urlcolor=blue}

\graphicspath{{images/}}

\DeclareMathOperator{\der}{d} 

\newcommand{\tff}{t_\mathrm{ff}}

\newcommand{\SigmaSFR}{\dot{\Sigma}_\mathrm{\star}}
\newcommand{\Sigmagas}{\Sigma_\mathrm{gas}}
\newcommand{\Sigmathr}{\Sigma_\mathrm{thr}}
\newcommand{\Sigmaavg}{\Sigma_\mathrm{avg}}
\newcommand{\Sigmamax}{\Sigma_\mathrm{max}}
\newcommand{\Sigmamin}{\Sigma_\mathrm{min}}
\newcommand{\YSO}{\mathrm{YSO}}

\newcommand{\eff}{\epsilon_\mathrm{ff}}
\newcommand{\SFR}{\dot{M}_\star}

\newcommand{\ndens}{\mathrm{cm^{-3}}} 

\newcommand{\pc}{\mathrm{pc}}
\newcommand{\MSun}{\mathrm{M}_\odot}

\newcommand{\Myr}{\mathrm{Myr}}

\newcommand{\coldensMPC}{\MSun\,\pc^{-2}}

\definecolor{lime}{HTML}{A6CE39}
\DeclareRobustCommand{\orcidicon}{%
        \begin{tikzpicture}
        \draw[lime, fill=lime] (0,0) 
        circle [radius=0.16] 
        node[white] {{\fontfamily{qag}\selectfont \tiny ID}};
        \draw[white, fill=white] (-0.0625,0.095) 
        circle [radius=0.007];
        \end{tikzpicture}
        \hspace{-2mm}
}

\newcommand{\orcidPS}{\href{https://orcid.org/0000-0001-7044-3809}{\orcidicon}}
\newcommand{\orcidAZ}{\href{https://orcid.org/0000-0001-9509-7316}{\orcidicon}}
\newcommand{\orcidPH}{\href{https://orcid.org/0000-0002-0472-7202}{\orcidicon}}
\newcommand{\orcidTC}{\href{https://orcid.org/0000-0002-2636-4377}{\orcidicon}}
\newcommand{\orcidAV}{\href{https://orcid.org/0000-0002-3320-7775}{\orcidicon}}
\newcommand{\orcidDR}{\href{https://orcid.org/0000-0001-5400-7214}{\orcidicon}}

\begin{document}
   \title{Stellar Feedback in the Star Formation-Gas Density Relation: Comparison between Simulations and Observations}

   \subtitle{}

   \author{P.~Suin
          \inst{1}\thanks{\email{paolo.suin@lam.fr}}\orcidPS
          \and
          A.~Zavagno \inst{1,2}\orcidAZ
        \and T.~Colman\inst{3}\orcidTC
        \and P.~Hennebelle\inst{3}\orcidPH
        \and A.~Verliat\inst{3}\orcidAV
        \and D.~Russeil\inst{1}\orcidDR
          }

   \institute{Aix Marseille Univ, CNRS, CNES, LAM Marseille, France 
   \and 
   Institut Universitaire de France, 1 rue Descartes, 75005 Paris, France 
   \and
   AIM, CEA, CNRS, Université Paris-Saclay, Université Paris Diderot, Sorbonne Paris Cité, F-91191 Gif-sur-Yvette, France           
             }

   \date{Received 21 July 2023 ; accepted 24 November 2023}

\authorrunning{P. Suin et al.}
\titlerunning{Stellar Feedback in the Star Formation-Gas Density Relation}

 
  \abstract
   {The impact of stellar feedback on the Kennicutt-Schmidt law (KS law), which relates star formation rate (SFR) to surface gas density, is a topic of ongoing debate. The interpretation of high-resolution observations of individual clouds is challenging due to the various processes at play simultaneously and inherent biases. Therefore, a numerical investigation is necessary to understand the role of stellar feedback and identify observable signatures.}
   {In this study, we investigate the role of stellar feedback on the KS law, aiming to identify distinct signatures that can be observed and analysed. By employing magneto-hydrodynamics (MHD) simulations of an isolated cloud, we specifically isolate the effects of high-mass star radiation feedback and protostellar jets.  
   Indeed, high-resolution numerical simulations provide a valuable tool to isolate the impact of stellar feedback on the star formation process, while also allowing us to assess how observational biases may affect the derived relation.}
   {We use high-resolution (<0.01~pc) MHD numerical simulations of a 10$^4\,\MSun$ cloud and follow its evolution, under different feedback prescriptions. The set of simulations contains four types of feedback: one with only protostellar jets, one with ionizing radiation from massive stars (>8~M$_{\odot}$), one with the combination of the two and one without any stellar feedback. 
   In order to compare these simulations with the existing observational results, we analyse their evolution by adopting the same techniques applied in observational studies. Then, we simulate how the same analyses would change if the data were affected by typical observational biases, including young stellar objects (YSOs) counting to estimate the SFR, limited resolution for the column density maps and a sensitivity threshold to detect faint embedded YSOs.
    }
   {Our analysis reveals that the presence of stellar feedback strongly influences the shape of the KS relation and the star formation efficiency per free-fall time ($\eff$).
   The impact of feedback on the relation is primarily governed by its influence on the cloud's structure. Additionally, the evolution of $\eff$ throughout the star formation event suggests that variations in this quantity can mask the impact of feedback in observational studies that do not account for the evolutionary stage of the clouds.
    Although the $\eff$ measured in our clouds results to be higher than what is usually observed in real clouds, upon applying prescriptions to mimic observational biases, we recover good agreement with the expected values. From that, we can infer that observations tend to underestimate the total SFR. Moreover, this likely indicates that the physics included in our simulations is sufficient to reproduce the basic mechanisms that contribute to set $\eff$.
   }
   {We show the interest of employing numerical simulations to address the impact of early feedback on star formation laws and to correctly interpret observational data. This study will be extended to other types of molecular clouds and ionizing stars, sampling different feedback strengths, to fully characterize the impact of \hii regions on star formation. }
   

   \maketitle
\section{Introduction}\label{intro}
\label{sec: intro}
The existence of a powerlaw relation between the star formation rate (SFR) per unit area and the gas surface density has long been established through observations of external Galaxies \citep{1959ApJ...129..243S, 1963ApJ...137..758S, 1989ApJ...344..685K}. The relation, known as the Kennicutt--Schmidt law (KS law), remarkably holds on scales spanning from entire galaxies to individual molecular clouds.

The dichotomy observed between starburst and non-starburst galaxies \citep{2004ApJ...606..271G, 2008AJ....136.2846B, 2012ApJ...745...69K}, might constitute a hint that stellar feedback intervenes in setting the star formation law \citep{2011ApJ...731...41O}, and \citet{2021ApJ...908...61K} suggested that this bimodal (or multimodal) relation could originate from changes in the small-scale structure of the molecular interstellar medium. Indeed, even nearby galaxies show variations on 1.5~kpc scales when observed at higher angular resolution \citep{2023ApJ...945L..19S}.

In effect, high-mass stars' feedback, such as photoionising radiation, stellar winds, and supernova explosions, greatly impact the surrounding environment \citep{2022MNRAS.509..272C}, but how these affect the star formation process is still not completely understood. 
For example, \hii regions are capable of blowing out the molecular cloud from which they were born. Yet, their expansion locally compresses the gas, enhancing its density and directly impacting the properties of new-forming clumps \citep{2020A&A...637A..40Z, 2021A&A...646A..25Z}. 
These events can affect scales from the tens of parsec down to sub-parsec scales, possibly altering the local star formation properties \citep{2017A&A...605A..35P,2021A&A...656A.101M}, or even promoting the formation of other massive stars by inhibiting gas fragmentation \citep{2022A&A...668A.147H}. 

Recently, results from the James Webb Space Telescope on the NGC~628 galaxy \citep{2023ApJ...944L..22B,2023MNRAS.521.5492M} shed light on the importance of understanding the role of early and smaller scales high-mass stellar feedback for a reliable description of star formation laws. However, effects acting on such small scales are invisible when averaged on galactic scales, and may reveal to be significant when considering singular molecular clouds. For this reason, observational studies have looked at the KS relation in several low- and high-mass molecular clouds \citep{2011ApJ...739...84G, 2015ApJ...809...87W, 2021ApJ...912L..19P, 2022ApJ...938..145B}, reporting a steepening of the exponent $\beta$ when passing from the largest galactic scales \citep[$\beta\approx1.5$;][]{2012ARA&A..50..531K, 2021ApJ...908...61K} to the smallest ones \citep[$\beta\approx2$;][]{2014ApJ...782..114E}.

Many theories attempt to explain the relation in all its shapes, but verifying or even constraining them represents a challenge for observational studies. Aside from resolution and sensitivity biases, the use of different tracers to derive the local physical conditions, and the multitude of processes ongoing simultaneously, make the comparison with the idealised cases extremely complicated. Moreover, these studies often rely on assumptions to compensate for the lack of three-dimensional information and uncertainties in theoretical models, which limits access to high-precision measurement of the real SFR--gas density relation. 

For instance, one of the most popular theories is the model proposed by \citet{2005ApJ...630..250K}. Under the assumption that the main force governing the process is gravity, the model proposes a linear dependence between $\SigmaSFR$ and $\Sigmagas$, with the typical timescale given by free-fall time $\tff$ of the region. In formula, this becomes
        \begin{equation}
            \SigmaSFR = \eff \frac{\Sigmagas}{\tff},
            \label{eq: modified KS}
        \end{equation}
where $\eff$, called star-formation efficiency per free-fall time, is the only parameter required. All the mechanisms other than gravity, feedback included, act to modify this coefficient. 
Despite its simplicity, the model yielded robust results \citep{2012ApJ...745...69K, 2015AJ....150..115U, 2018ApJ...861L..18U}, opening the path to more refined theories that can account for diverse local environmental properties. In particular, models incorporating the concept of multi-free-fall time are particularly promising \citep{2011ApJ...743L..29H, 2011ApJ...730...40P, 2012ApJ...761..156F, 2018ApJ...863..118B}. Unlike the original model, which relies on average density, these newer theories infer a density distribution to more accurately describe the gaseous environment. These models not only align more closely with observational data, but they also help to explain the higher rates of star formation observed in starburst galaxies \citep{2013MNRAS.436.3167F}.
Nevertheless, although elegant from an analytical perspective, using the local free-fall time represents a major challenge for astronomers. Specifically, obtaining this quantity typically requires resorting to strong assumptions to estimate the average volume density, such as the uniform-sphere approximation.

Even deriving the local SFR is complicated. 
In recent works conducted on nearby Galactic clouds, this is inferred by counting the number of detected young stellar objects (YSOs), but their mass and age had to be assumed \citep[e.g.][]{2013A&A...559A..90L, 2017A&A...606A.100L}. Additionally, in the densest regions of clouds, the count of YSOs is bound by the instruments' sensitivity \citep{2016AJ....151....5M}. All these problems might explain why some studies claim that this parameter remains constant at all column densities \citep{2021ApJ...912L..19P}, while others even suggest that the model cannot provide a satisfactory description of the process in single molecular clouds \citep{2014ApJ...782..114E, 2016ApJ...833..229L, 2022ApJ...938..145B}

Part of the problem could be that the molecular clouds used in these studies vary in mass, structure, star formation activity and evolutionary stage. Some contain high-mass stars, while others are known to be low-mass star-forming regions. Encompassing all these effects simultaneously, the individual impacts on the star formation laws blur together. If stellar feedback actually plays a role, accurately separating the clouds according to their physical condition could solve the tension.

To overcome these challenges, we employed high-resolution numerical simulations to study the possible impact of different types of stellar feedback on the KS relation across the temporal evolution of one cloud. 
Using simulations allowed us to investigate the time evolution without having to deal with many of the difficulties associated with observations. Our simulation suite consisted of two runs where either protostellar jets or ionizing radiation is modelled, one run in which both feedback mechanisms are included, as well as one simulation without any stellar feedback. In this way, we could analyse the effect of each of these feedback mechanisms separately.  

The primary objectives of our work are twofold. First, we aimed to characterize the impact of stellar feedback on the star-formation laws at the cloud scale and to disentangle the temporal evolution of these effects. Second, we sought to identify observational signatures associated with these phenomena. To achieve this, we analysed the simulations using observational techniques and compared our findings with recent observational results. Additionally, we considered the biases that may affect these procedures, focusing on the impact of YSOs counting to estimate the SFR, low resolution on the column density maps and sensitivity limit to detect faint and extinct stars. The purpose of our investigation is to improve our understanding of the intricate relationship between feedback, evolutionary stage, and the observed properties of star-forming clouds.

The paper is organised as follows. In Section~\ref{simu} we describe the simulations, and in Section~\ref{methods} we explain the methods adopted to derive the KS relation and $\eff$. Section~\ref{res} contains the results of the analysis, while in Section~\ref{sec: biases} we investigate the possible impact of observational biases on the observed relations. Finally, in Section~\ref{dis} we contextualise the obtained results from a physical point of view. Conclusions are given in Section~\ref{conc}.

\section{Numerical simulations} \label{simu}
In our study, we analysed the simulations presented in \citet[][]{2022A&A...663A...6V}. 
This set of four runs followed the evolution of the same isolated $10^4\,\MSun$ cloud under the influence of diverse feedback prescriptions: one with both ionising radiation and protostellar jets,  one with only ionising radiation, one with only protostellar jets, and one without any feedback. For simplicity, we will refer to these runs as \texttt{HIIR+PSJ}, \texttt{HIIR}, \texttt{PSJ} and \texttt{NF}. Thanks to the wide range of physical processes implemented and the high resolution achieved (order of thousands of AU), these simulations provide an ideal test-bed for investigating the effect of stellar feedback on the star-formation properties and for testing the observational techniques employed in the field. 
Here we briefly describe the global properties and the setup of the model, referring to \citet{2022A&A...663A...6V} for a complete description of how the various physical processes are implemented.

\subsection{Code and numerical parameters}
The magnetohydrodynamic simulations were carried out using the adaptive mesh refinement (AMR) code \texttt{RAMSES} \citep{2002A&A...385..337T}. The domain was set up as a cubic box of $L=30.4\,\pc$, with open boundaries and an initial resolution of $128^3$ cells. Adopting five AMR levels, the maximum resolution achieved in each simulation is $7.4\cdot10^{-3}\,\pc$ $(1.5\cdot10^3\,\mathrm{AU})$. The refinement criterion assures that the local Jeans length is always resolved with at least 40 cells. When the density of a cell at the highest AMR level exceeds $10^7\,\ndens$, a sink particle is created \citep{2014MNRAS.445.4015B}. These sinks interact with the environment by accreting material from their surroundings and ejecting 1/3 of it under the form of protostellar jets. Each time the sinks accrete $120\,\MSun$ of gas overall, an ionising star is created inside one of them. The stellar mass is drawn from a \citet{1955ApJ...121..161S} distribution between 8 and $120\,\MSun$. Then, the parametric mass-UV luminosity relation given in \citet{1996ApJ...460..914V} sets its ionising flux. The sequence of stellar masses extracted had been preserved between simulations \texttt{HIIR} and \texttt{HIIR+PSJ} so that they formed the same stars throughout the cloud's evolution.

\subsection{Initial conditions}
Initially, the cloud was modelled as an approximated $10^4\,\MSun$ Bonnor--Ebert sphere, with a diameter of $15.2\,\pc$ and a scale radius $r_0=2.5\,\pc$. Its density was distributed according to
\begin{equation}
    n(r) = \frac{n_0}{1+\left(\frac{r}{r_0}\right)^2}
    \label{eq: cloud density distribution}
\end{equation}
with $r$ the distance from the box centre, and the central density $n_0=800\,\ndens$. The rest of the cube is filled with a uniform density of $8\,\ndens$ up to $r=L$, and the density in the remaining corner regions is set to $1\,\ndens$.

The gas thermal behaviour was parameterised with a cooling function that takes into account several chemical processes (see \citealt{2005A&A...433....1A} for a detailed description). This set the initial temperature of the medium at 10\,K in the densest parts. 
The magnetic field was set to have a uniform mass-to-flux ratio of 8 in the $x$ direction. 
Finally, the turbulence had been initialised with a velocity field normalised to have a Mach number of 6.7 in the central region, compatible with what is observed in Galactic star-forming clouds \citep{2022ApJS..262...16M}. 
A Kolmogorov power spectrum is then imposed to roughly reproduce the behaviour of the low-density gas filling the edges of the box, where the high temperatures make the flow sub/transonic and approximately incompressible \citep[e.g.][]{1994PhFl....6.2133P}.
An alternative could have been to favour the low-temperature gas at the centre of the domain choosing a Burgers power spectrum, which more suits its high-Mach regime \citep{2000nlin.....12033F}. In Section\,\ref{sec: eff} we qualitatively discuss how this choice would not change the results of this work.
However, we highlight that, given the idealised setup, the specific choice of the power spectrum has a marginal impact on the accuracy of the initial conditions. In effect, the mixture of high- and low-temperature gas is such that a simple powerlaw power spectrum cannot reproduce the turbulence spectrum characteristic of the medium\footnote{
It's worth noting that modern simulations can overcome this limitation by using zoom-in techniques, modelling larger galactic environments and then narrowing the focus to specific regions, using them as initial set-up for more detailed simulations \citep{2015MNRAS.447.3390D, 2017MNRAS.472.4797S}. While yielding greater reliability, this approach sacrifices generalisability, as the results become are directly linked to the unique properties of the selected region.
}. Since the turbulent velocity field in our simulation is only an initial condition and not continuously driven, the velocity power spectrum will naturally evolve to adopt the energy cascade that is physically consistent with the system.

To conclude, we can summarise the initial parameters in terms of the characteristic free-fall timescales of the central region, $\tff\approx1.5\,\Myr$\footnote{Obtained using a mean molecular weight of $1.4$, as required by the parametric cooling function.}. With these initial conditions, the sound-crossing timescale of the inner region inside $r_0$ is defined by $\frac{\tff}{t_\mathrm{sc}}=0.15$, while the Alfvén-crossing time by $\frac{\tff}{t_\mathrm{ac}}=0.2\,\tff$. The turbulent-crossing time was set equal to the free-fall time.

\subsection{Evolution}

\begin{figure}[t]
    \centering
     \includegraphics[width=1\linewidth]{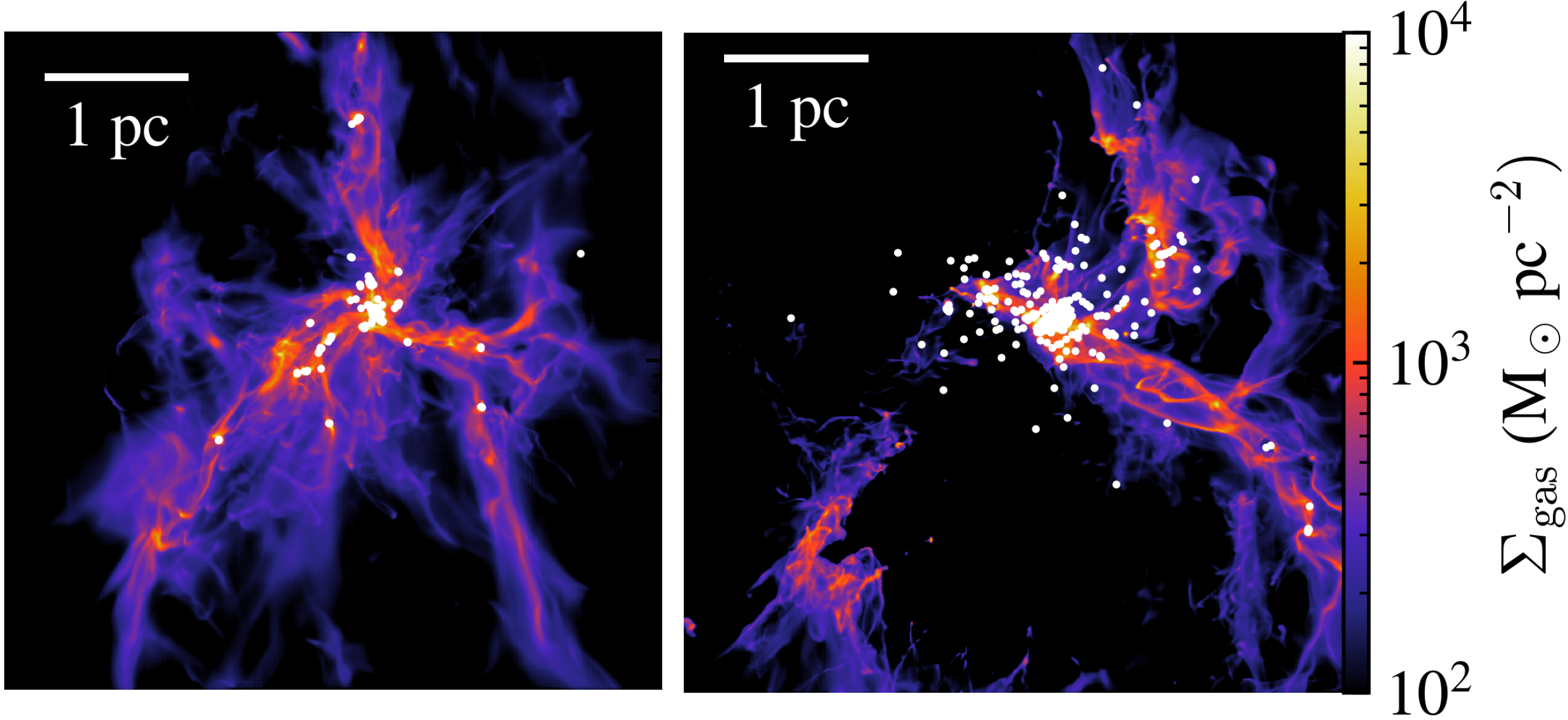}
    \caption{Density projections along the $z$-axis for two snapshots of simulation \texttt{HIIR+PSJ}, at 2.3\,Myr (left) and 3.15\,Myr (right). White dots mark the sink particles' positions. The first captures the formation of the central hub, with three filaments converging into it. In the second we can see how the giant \hii region, originating from the lower left filament, restructured the cloud after about $0.4\,\Myr$. }
    \label{fig: simulations}
\end{figure}
\begin{figure}
    \centering
     
     \includegraphics[width=1\linewidth]{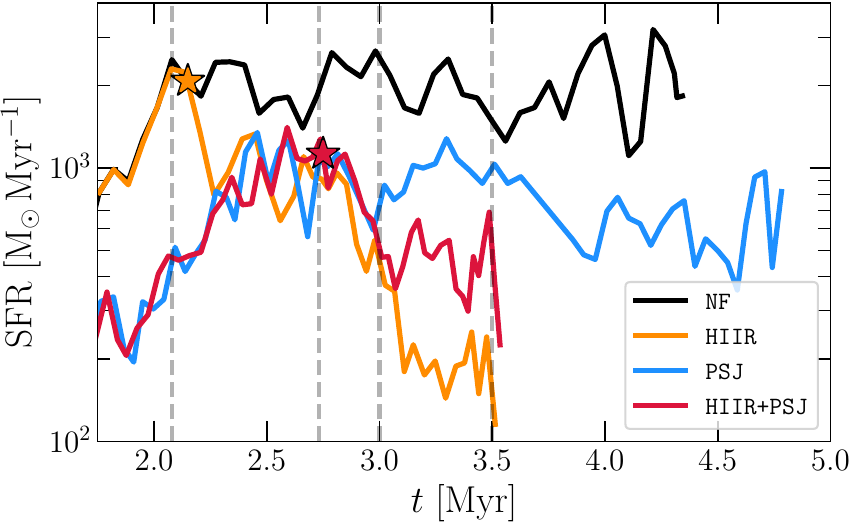}
    \caption{Temporal evolution of the SFR in the different simulations. The black line shows the simulation without any feedback. The run with only jet and the one with only ionising radiation are represented in blue and orange, respectively. Finally, the red line individuates the simulation with both feedback active. The stars mark the formation of the giant \hii region. The vertical dashed lines highlight the approximate time of the snapshots used in Figures~\ref{fig: SFR evolution}-\ref{fig: eff evolution}. }
    \label{fig: SFR glob}
\end{figure}

Given the initial conditions, that initialised the isolated cloud with a central density enhancement, the system naturally developed a hub system within a couple of Myr, with three filaments converging onto it (cf. Figure~\ref{fig: simulations} left). The four simulations evolved identically until the formation of the first star at $1.45\,\Myr$. Afterwards, jets started to reshape the cloud and runs \texttt{PSJ} and \texttt{HIIR+PSJ} departed from the ones without jets. Lacking any feedback initially, the \texttt{HIIR} and \texttt{NF} clouds rapidly increased the total SFR -- measured as the total amount of gas converted into sinks throughout two subsequent snapshots, and divided by their temporal separation (generally $\approx0.04\Myr$). In those simulations the SFR reached its maximum at $\approx2\,\Myr$ (Figure~\ref{fig: SFR glob}), while in the other two, the presence of protostellar jets lowers the SFR by a factor of four initially\footnote{A similar impact of jets on the total SFR has also been observed in similar studies, like \citet{2015MNRAS.450.4035F} - who reports a reduction of roughly a factor of three -, \citet{2023ApJ...954...93A} - roughly 2 -, and \citet{2018MNRAS.475.1023M} - also slightly more than two. Although the ratio between the SFRs in \texttt{NF} and \texttt{PSJ} simulations is initially higher, agreement with these studies is recovered once both simulations reach the steady-state.}. Between $1.8\,\Myr$ and $2.15\,\Myr$, a few ``low'' mass ionising stars (between $8$ and $18\,\MSun$) formed in \texttt{HIIR}, but their associated \hii regions only mildly affected the global evolution. By $2.15\,\Myr$, while the gravity-only run settled down to a roughly constant $\SFR^\mathrm{tot}$, the onset of a powerful \hii region, generated by a $100\,\MSun$ star, caused the SFR to drop at the same level of the other two feedback simulations. In a little more than $1\,\Myr$ the cloud was wiped out and the SFR flattened at low values. In \texttt{HIIR+PSJ}, the same ionising stars formed at $\approx2.75\,\Myr$. Afterwards, the behaviour mirrored that of radiation-only simulation. In the right graph of Figure~\ref{fig: simulations} we can see the \hii region in the process of dispersing the gas, although some triggered overdensity regions clearly arose from the shell compression. Instead, the SFR of \texttt{PSJ}, as that of \texttt{NF}, remained roughly constant up to the end of the simulation. 

Therefore, it is clear that the presence of ionising stars overall reduces $\SFR^\mathrm{tot}$. Nevertheless, in literature, it is still greatly debated whether or not high-energy radiation can enhance the SFR and modify the SF properties on a local scale \citep{2009A&A...494..987P, 2011EAS....51...45E, 2013MNRAS.431.1062D, 
2013A&A...554A...6R,
2020MNRAS.493.4643M, 2020ApJ...904..192W}. 
In this paper, we approached this issue from an observational perspective, investigating the underlying physical mechanisms and identifying potential biases arising from the absence of a third dimension and temporal evolution.

\section{Methods} \label{methods}
\subsection{Extraction of quantities}
\begin{figure}
    \centering
\includegraphics[width=1\linewidth]{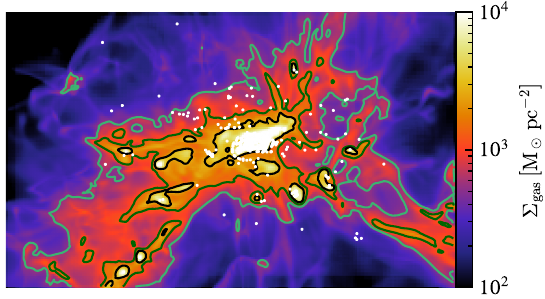}
    \caption{Zoom of the central cluster of \texttt{HIIR+PSJ} simulation, at $2.7\,\Myr$. Light green, dark green and black lines mark the contours at 500, 2500, 5000$\,\coldensMPC$ respectively. The region covers an area $1.8\,\pc\times1.1\,\pc$.}
    \label{fig: contours}
\end{figure}

To extract the relation between the SFR and the gas density we first projected the simulation's snapshots along the three axes at the highest resolution, giving maps of $4096^2$ pixels.
Then, we divided the column density map into a set of 500 contours, evenly log-spaced between $80\,\coldensMPC$ and $\Sigmagas=3\cdot10^4\coldensMPC$. As in some recent observational works \citep{2021ApJ...912L..19P, 2022ApJ...938..145B, 2022MNRAS.511.1431H}, the properties of a given contour level (the average density $\Sigmaavg$ and its area $A$) were obtained integrating over all the pixels above the corresponding threshold $\Sigmathr$. Figure~\ref{fig: contours} gives an idea of the contours' shape, highlighting three levels at 500, 2500, 5000$\,\coldensMPC$ in simulation \texttt{HIIR+PSJ} at $2.7\,\Myr$. 
We did not make distinctions between detached regions that may have formed while increasing $\Sigmathr$, but we averaged all the quantities on the contour total area. Following \citet{2022ApJ...938..145B}, we refer to this approach as \textit{cumulative}, to avoid any confusion with the \textit{differential} one, where quantities are extracted from the region included in between two subsequent contours. We retained only contours with a total surface area greater than 500 pixels, to exclude possible resolution effects on the data. As shown in Section~\ref{subsec: resolution} and Appendix~\ref{app: 3d-2d}, this criterion resulted to be conservative in most of the cases.

Finally, we computed the SFR density $\SigmaSFR$ associated with each contour as
\begin{equation}
    \SigmaSFR=\frac{\SFR}{A},
    \label{eq: dependences SigmaSFR}
\end{equation}
where $\SFR$ is the total mass accreted per unit time by the sinks enclosed in the contour, calculated with the same method described in the previous Section. We remark that observational studies do not have direct access to such quantity. Indeed, the limited knowledge of YSOs' evolution leads to significant uncertainties both on their age and mass. The common approach is to assume an average mass and age for all of them, so that the SFR is proportional to the number of YSOs counted within a contour \citep[e.g.,][]{2014A&A...566A..45L, 2016A&A...587A.106Z, 2017A&A...606A.100L, 2021ApJ...912L..19P}. 
However, adopting this approach in our analysis would probably be affected by numerical effects. Indeed, despite the high resolution, it is not possible to carefully follow the formation of single stars and therefore to reproduce the correct initial mass function (IMF), because this would require resolution of the order of tens of AU \citep{2018A&A...611A..89L}. Presently, no simulation has been capable of covering such a wide range of scales while including all the physical processes treated here. We will further address this issue in Section~\ref{subsec: yso counts}.

\subsection{Free-fall timescale}
The model introduced by \citet{2005ApJ...630..250K} has the advantage of providing a physical interpretation of the law, obtaining an exponent of 1.5, similar to what is observed at galactic scales (in 2D $\tff\sim\Sigmagas^{-0.5}$). However, the drawback is the introduction of the new variable $\tff$, particularly difficult to obtain from cloud-scale observations. The common approach is to evaluate $\tff$ from the projected quantities under uniform sphere approximation \citep{2011ApJ...729..133M, 2021ApJ...912L..19P, 2022ApJ...938..145B}. Since our purpose is to supply a possible numerical counterpart to observational studies of the KS law, we employed here the same technique. In particular, for each contour, we extracted its typical free-fall timescale estimating its volume density as 
        \begin{equation}
            \rho_\mathrm{avg}= \frac{3\sqrt{\pi}}{4} \Sigmaavg A^{-0.5},
            \label{eq: spherical approx}
        \end{equation}
and thus computing $\tff = \sqrt{3\pi / 32 G \rho_\mathrm{avg}}$, with $G$ the gravitational constant.

Studies conducted on galactic surveys frequently report low values of the $\eff$, ranging around 0.01  \citep{2008AJ....136.2782L, 2017ApJ...846...71L, 2018ApJ...861L..18U}. However, recent works dealing with Milky Way clouds, instead, report higher values, but with a significant dispersion, going from $\eff<10^{-3}$ to $\eff>0.1$, with peaks of 0.3 \citep{2011ApJ...729..133M, 2016ApJ...833..229L}. 

As mentioned in Section\,\ref{sec: intro}, more evolved models were developed in the past decade. Computing a typical timescale from the local properties of the region, theories using the multi-free-fall time proved to provide valid results on galactic and extragalactic scales. However, we highlight that our intent is to compare the SF properties of the inner structures of molecular clouds as recovered from numerical simulations and observations. In this context, we are not aiming to utilise the most advanced model to calculate $\eff$, but rather follow the very same approach used in the works based on observation we are comparing to.

\section{Resulting star-formation relations} \label{res}
\subsection{Kennicutt-Schmidt relation}

\begin{figure*}
    \centering
    \includegraphics[width=1\linewidth]{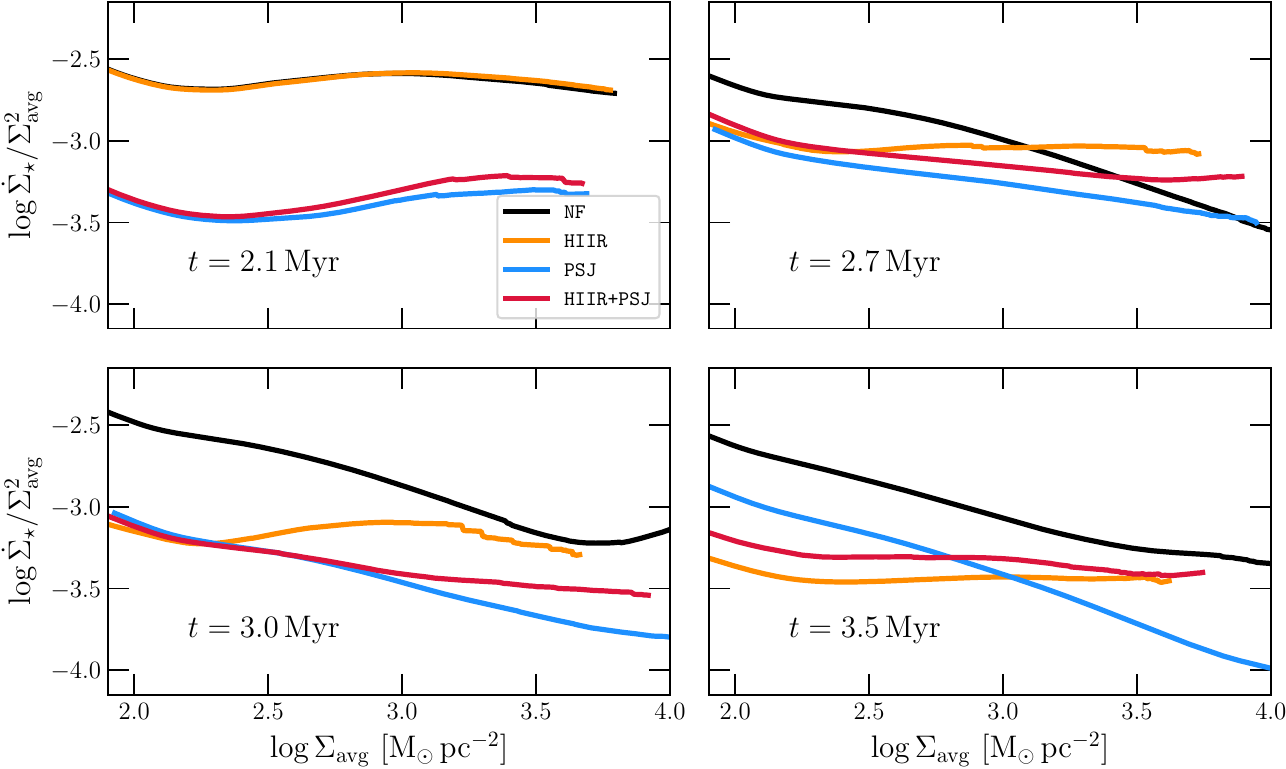}
    \caption{
    Comparison of the star formation--gas density relation between the different simulations, shown at four different times ($t=2.1,\,2.7,\,3.0,\,3.5\,\Myr$), that illustrates the evolution of the cloud. In \texttt{HIIR} the giant \hii region develops after $2.2\,\Myr$, while for \texttt{HIIR+PSJ} it starts to restructure the environment at later times ($>2.75\,\Myr$). The colour code is the same as in Figure~\ref{fig: SFR glob}.
    \label{fig: SFR evolution}}
\end{figure*}
We show the resulting KS relation obtained for our clouds in Figure~\ref{fig: SFR evolution}. 
To facilitate the comparison between simulations, we scaled the $y$-axis by $\Sigmaavg^2$, which represents the characteristic slope observed in the radiation simulations. This is also consistent with typical values reported in observational studies conducted on cloud scales \citep[see e.g.][]{2011ApJ...739...84G, 2013ApJ...778..133L, 2021ApJ...912L..19P}. The plots show the projection along the $z$-axis, but comparisons with other lines of sight yielded similar conclusions (see Appendix \ref{app: los}, \cite{2019MNRAS.488.1407K}).
We selected four specific snapshots to represent key stages in the simulations' evolution:
\begin{itemize}
\item at $2.1\,\Myr$. By this time the SFR of run \texttt{NF} has almost reached the steady state (Figure~\ref{fig: SFR glob}), while cloud \texttt{HIIR} has not been affected by the eruption of the large \hii region;
\item at $2.7\,\Myr$. Here, due to the presence of the $100\,\MSun$ star, the SFR of the radiation-only simulation decreased and reached the level of the runs with jets. The same star has yet to be formed in \texttt{HIIR+PSJ};
\item at $3.0\,\Myr$. At this time, the four clouds all present significant differences between each other, and the main $\hii$ region has started to restructure the full-feedback run (\texttt{HIIR+PSJ});
\item finally, we plot the situation at the very last moments of the radiation simulations, at $3.5\,\Myr$.
\end{itemize}

The plots reveal that the different feedback prescriptions lead to significant differences in the clouds' evolution. In the early stages, when only a few small \hii regions developed, jets dominate as the primary feedback mechanism. Therefore, at $2.1\,\Myr$, we observe strong similarities between the runs \texttt{NF} and \texttt{HIIR} (without jets) and \texttt{PSJ} and \texttt{HIIR+PSJ} (with jets). Similarly to what is displayed in Figure~\ref{fig: SFR glob}, we see that stellar ejecta reduce the global SFR by a factor of four\footnote{The very left-hand side corresponds to the widest contours, which include all the others. From those we can recover the average behaviour of the cloud.}$^,\,$\footnote{Notice that this factor is greater than the loss of accretion due to the expulsion of material, which alone contributes to decreasing the total SFR by a factor of 1.5.}, without sensibly modifying the shape of the relation.

By $2.7\,\Myr$ the situation has changed substantially. Overall, the KS relation has become an almost pure powerlaw in all simulations. Toward the low-density edge, a clear change of slope is visible in runs with feedback, while it is almost invisible in run \texttt{NF}. Probably, this broken powerlaw identifies the point at which stellar feedback counteracts the global collapse, and the cloud's KS relation departs from that typical of a free-fall motion. Moreover, we can note \texttt{HIIR+PSJ} showing a slightly steeper relation compared to the jets-only simulation. Interpreting the graph, this means that the relative fraction of gas mass that is converted into stars per unit time increases at high densities when ionising radiation is included, even though at this time the total SFR is similar (Figure~\ref{fig: SFR glob}). This behaviour is even more accentuated comparing runs \texttt{HIIR} and \texttt{NF}, where the first achieves higher values of $\SigmaSFR$ at high densities, with an exponent of its KS relation of $\approx 2$.

At $3.0\,\Myr$, the different exponent of the \texttt{HIIR+PSJ} and \texttt{PSJ} KS relation became even more evident. Another visible feature is the departure from the powerlaw behaviour of run \texttt{HIIR} at $\Sigmaavg\approx10^3\,\coldensMPC$. Given the sudden decline, we can deduce that the main \hii region, dispersing the gas, sensibly lowered the column density around a sink formation site. Then, when $\Sigmathr$ exceeds the maximum column density of that region, the observed SFR suddenly drops. 
Run \texttt{NF}, instead, exhibited an increase in $\SigmaSFR$ toward higher column densities. In the simulation, this increase coincides with the moment in which a clump, formed in one of the filaments, converges into the central hub. 

Finally, at $3.5\,\Myr$, the large \hii region has fully developed in the simulation \texttt{HIIR+PSJ} and its $\SigmaSFR$ versus $\Sigmaavg$ relation mirrors that of the radiation-only simulation. Its slope, as well as the one from \texttt{HIIR}, closely approaches $\Sigmaavg^2$. With the system reaching a steady state, the \texttt{NF} run does not display significant differences compared to the previous graph, while \texttt{PSJ} experiences a further steepening of the relation.

\subsection{Modified Kennicutt-Schmidt relation}
\label{sec: eff theory}
\begin{figure*}
    \includegraphics[width=1\linewidth]{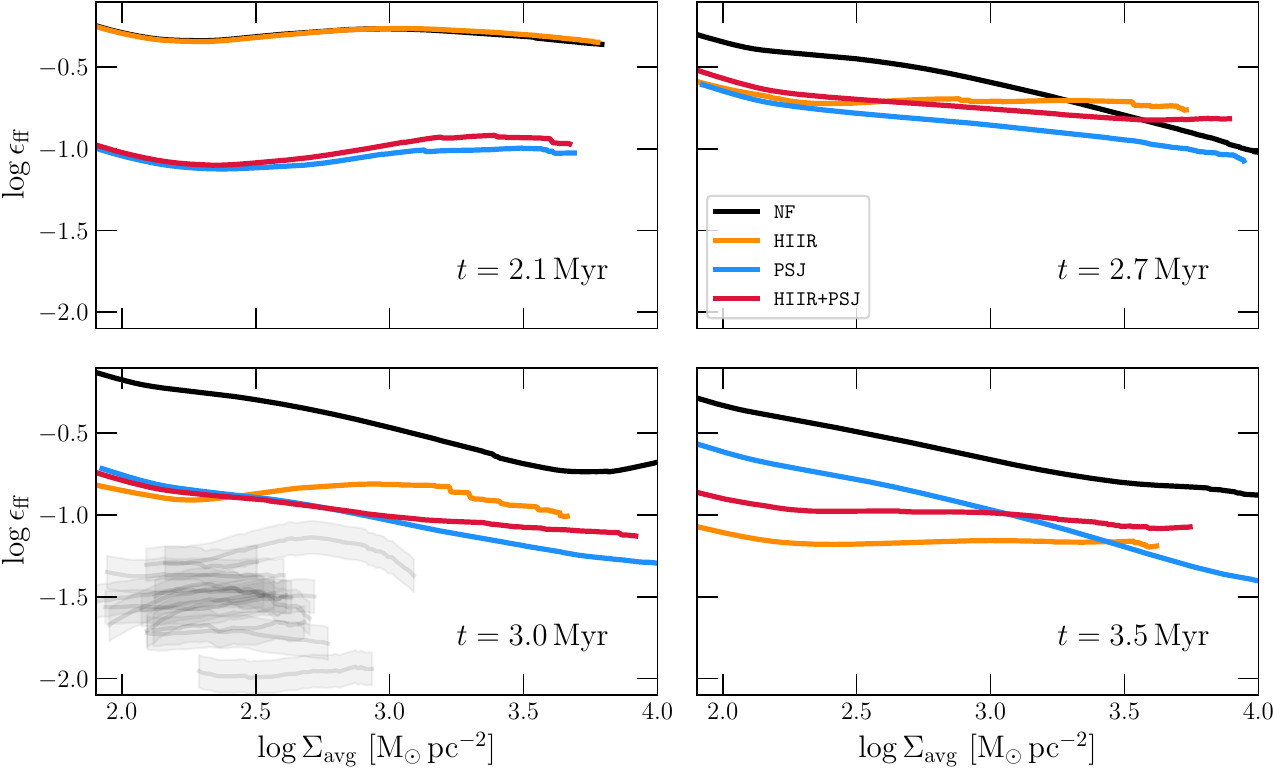}
    \caption{Evolution of the modified KS relation. The colour code is the same as in Figure~\ref{fig: SFR glob}. In the graph at $3\,\Myr$, we report the values of $\eff$ of the 12 Galactic clouds studied in \citet{2021ApJ...912L..19P}, excluding contours with $\Sigmaavg$ above their reliability criterion (grey lines). The grey shaded areas represent the uncertainties they gave to the $\eff$ measures. The axes ratios are kept the same as in Figure~\ref{fig: SFR evolution}, to aid a visual comparison of slopes with the respective counterparts in the KS relation.
    \label{fig: eff evolution}}
\end{figure*}
\label{sec: eff}
\begin{figure}
    \includegraphics[width=1\linewidth]{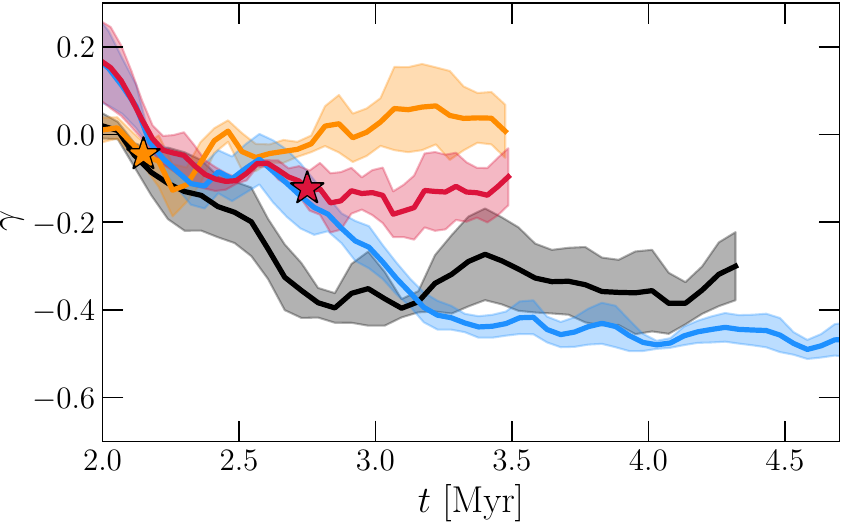}
    \caption{Evolution of the slope of $\eff$ across the simulations. The shaded areas outline the $1\,\sigma$ standard deviation of the fitted exponent as obtained from the three different axes of projection. The stars mark the formation of the main \hii region, as in Figure~\ref{fig: SFR glob}
    \label{fig: eff slope}}
\end{figure}

The evolution of $\eff$ across time (Figure~\ref{fig: eff evolution}), obtained inverting Eq.\,\ref{eq: modified KS}, closely reproduces that seen in Figure~\ref{fig: SFR evolution}. However, the differences in both slope and scatter between the \texttt{PSJ} and radiative simulations are less pronounced. In effect, one of the major strengths of the modified relation is its reduced scatter among different objects, compared to the traditional KS law \citep{2022MNRAS.511.1431H}. Observational datasets typically consist of multiple clouds at various evolutionary stages, characterised by different strengths of feedback mechanisms within the complex. Given that our simulated clouds feature different feedback prescriptions, the smaller dispersion observed in the four $\eff$ values aligns well with expectations based on observational results.

As in Figure~\ref{fig: SFR evolution}, the curves from radiative simulations tend to achieve higher values toward the high-density end of the graphs. Additionally, the $\eff$ of these simulations appears almost flat, while in runs \texttt{NF} and \texttt{PSJ} it shows a steady decline of $\eff$ as the density increases. 
Indeed, this qualitative result is confirmed when fitting a powerlaw like
            \begin{equation}
                \eff \sim \Sigmaavg^{\gamma}.
            \end{equation}
At each snapshot, we fitted\footnote{We employed the \texttt{Python} built-in function \texttt{curve\_fit}} the relation using only contours that satisfy the 500 pixels contours and have $\Sigmaavg>300\,\coldensMPC$. This allowed us to exclude the different relation visible at lower densities -- i.e. larger scales --, which might be affected by the initial condition of the cloud. Indeed, the evolutionary timescales of these regions are much longer, and the system employs more time to relax from the initial conditions here. We repeated this procedure for all three projection axes and assigned to the average $\gamma$ an uncertainty given by the standard deviation from the three lines of sight (the errors from the fit procedure were completely negligible).

Figure~\ref{fig: eff slope} shows the evolution of the exponent $\gamma$ across the simulation. Again, we see that only simulations with radiation ultimately set on $\gamma$ values close to zero. A $\eff$ independent of $\Sigmaavg$ has also been reported in nearby Galactic cloud by \citet{2021ApJ...912L..19P}, although in their dataset the value of $\gamma$ scatters between -0.33 (AFGL 490) and 0.06 (Cygnus-X)\footnote{This can be recovered from the second model fitted in Table 1 (sixth column), since $\gamma = a-1$.}.

Nevertheless, compared to their work, we found sensibly higher values of $\eff$. 
For comparison, we overplotted the curves they obtained for the 12 Galactic clouds in the graph at $3\,\Myr$ in Figure~\ref{fig: eff evolution}, and we can see how the two measures differ by a factor of $\approx5$.
While the influence of the SFR prescription cannot be ruled out entirely, it is improbable that such a difference is attributable \textit{entirely} to the code setup\footnote{
Our findings are not the first to reveal a mismatch between the observed and simulated SFRs and $\eff$. Specifically, \citet{2015MNRAS.450.4035F} has also highlighted that numerical simulations often result in higher SFRs than those reported in observational studies.
}.   Indeed, we expect the total simulated SFR to be reasonably accurate in runs with jets\footnote{The lack of gas expulsion from sinks in \texttt{HIIR} and \texttt{NF} causes an overestimate of the star formation rate by a factor of 1.5.}, and the sink spatial distribution reliable at least up to $\Sigmathr>3000\,\coldensMPC$ (roughly 3.5 in the graph, cf.\,with Figure~\ref{fig: Cloud Structure}). Moreover, even though our simulations could not accurately follow the dynamical evolution of sinks due to the softening of gravity, \citet{2013A&A...559A..90L} and \citet{2013ApJ...778..133L} showed in real clouds that the diffusion process due to $N$-body interactions does not contribute significantly to the observed $\SigmaSFR$ at low densities. 
Finally, we outline that adopting a Burgers power spectrum as initial condition for turbulence would probably not solve the tension observed with the $\eff$ reported in \citet{2021ApJ...912L..19P}. Indeed, a steeper power spectrum means that much less kinetic energy is stored in small scales, leading to more coherent large-scale flows. Therefore, imposing a Burgers spectrum in the velocity field would have enhanced the formation of bound structures \citep{2005MNRAS.361....2C}, likely increasing the cloud's SFR and consequently the discrepancy with the SFRs obtained from the observational data.

Therefore, we looked at three potential biases that may influence observational measurements. We emphasise right away that our intent is to understand how different biases could interact with the feedback mechanisms to affect the results. Although full agreement with \citet{2021ApJ...912L..19P} $\eff$ measures is recovered, giving a precise quantitative prediction of the impact of these biases is beyond the purpose of this paper. 

\section{Impact of biases on the observed relation} 
\label{sec: biases}
\begin{figure*}[h!]
    \includegraphics[width=1\linewidth]{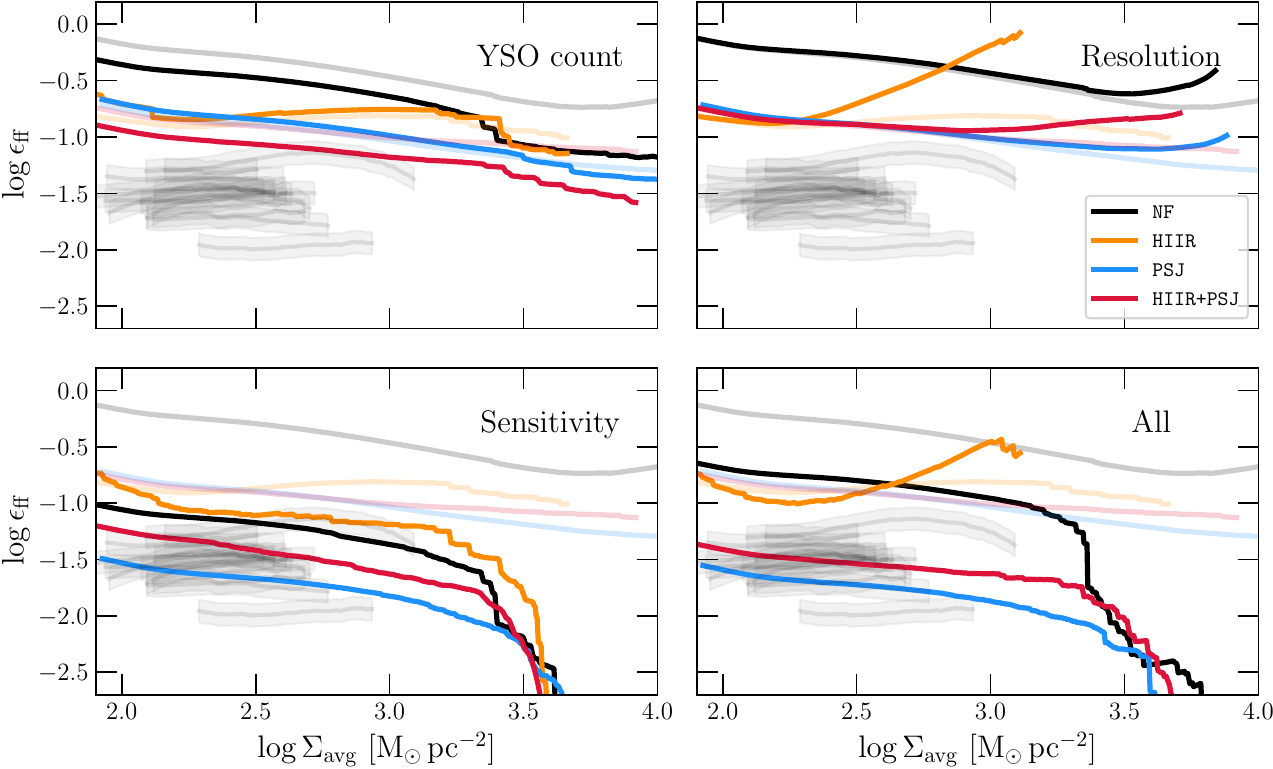}
    \caption{Effect of four different observational biases on the $\eff$--$\Sigmaavg$ relation of the clouds at $3\,\Myr$. The blended curves are the same as panel $3\,\Myr$ in Figure~\ref{fig: eff evolution}, while the grey lines are the 12 Galactic clouds studied in \citet{2021ApJ...912L..19P}.
    \label{fig: biases}}
\end{figure*}
We examined, in our simulations, the effect of three biases that affect observations: the assumption that the SFR is proportional to the number of Class~0/I YSOs detected, the impact of resolution on column density maps, and the presence of embedded stars not visible to the telescope due to extinction. In this Section, we aim to understand their impact on the observed $\eff$--$\Sigmaavg$ relation. In Figure~\ref{fig: biases}, we plotted the effects of these biases separately, using the snapshot at $3\,\Myr$ as a reference (blended curves in the background). At that time, the contribution of ionising radiation in \texttt{HIIR+PSJ} was not as dominant as in \texttt{HIIR} yet. This allowed us to study how these biases intertwine with the different levels of feedback.

\subsection{YSO counts}
\label{subsec: yso counts}
The first bias we considered is caused by the uncertainty on detected YSOs' age and mass. Since observations cannot access the temporal evolution of the star-forming clouds, observers rely on theoretical models to constrain these properties in young stars. As mentioned in Section \ref{methods}, the usual procedure is to assign a mean mass and age to each of them. The mass corresponds to the average mass of the IMF, which is typically around $0.5\,\MSun$. For Class~0/I objects, the age is commonly assumed to be $0.5\,\Myr$ \citep{2015ApJS..220...11D, 2022PASP..134d2001M}. With these values, the SFR of each contour is $\SFR=1\,\MSun\,\Myr^{-1} \times N_\YSO$, where $N_\YSO$ is the number of young stars it contains. Imitating this technique, we counted only the sinks younger than $0.5\,\Myr$. However, we explained in Section~\ref{methods} that the IMF in our cloud is not well sampled due to insufficient resolution. Indeed, the average stellar mass resulted to be higher than the expected $0.5\,\MSun$, with $\langle m_\mathrm{sink}\rangle\approx0.75\MSun$ in the feedback runs and $2\,\MSun$ in \texttt{NF} at the end of the simulation. Using a coefficient of $1\,\MSun\,\Myr^{-1}$ in the above equation would underestimate the total SFR by a factor of $\langle m_\mathrm{sink}\rangle/0.5\,\MSun$. Therefore, we corrected this by assigning to each contour an internal SFR given by $\SFR=\langle m_\mathrm{sink}\rangle/(0.5\,\Myr) \times N_\YSO$. 

The first plot in Figure~\ref{fig: biases} (upper left) demonstrates that this particular approximation well reproduces the true SFR of the cloud. 
Except for \texttt{NF}, the detected $\eff(\Sigmaavg)$ preserves good agreement at all contours. Minor deviations are observed at high-density contours, probably due to random fluctuations in the instantaneous SFR. In these regions, $\tff$ becomes much smaller than the assumed build-up time of $0.5\,\Myr$, so that the integration time becomes much longer than the dynamical timescale \citep{2021ApJ...912L..19P}.

\subsection{Spatial resolution}
\label{subsec: resolution}
\begin{figure}
    \includegraphics[width=1\linewidth]{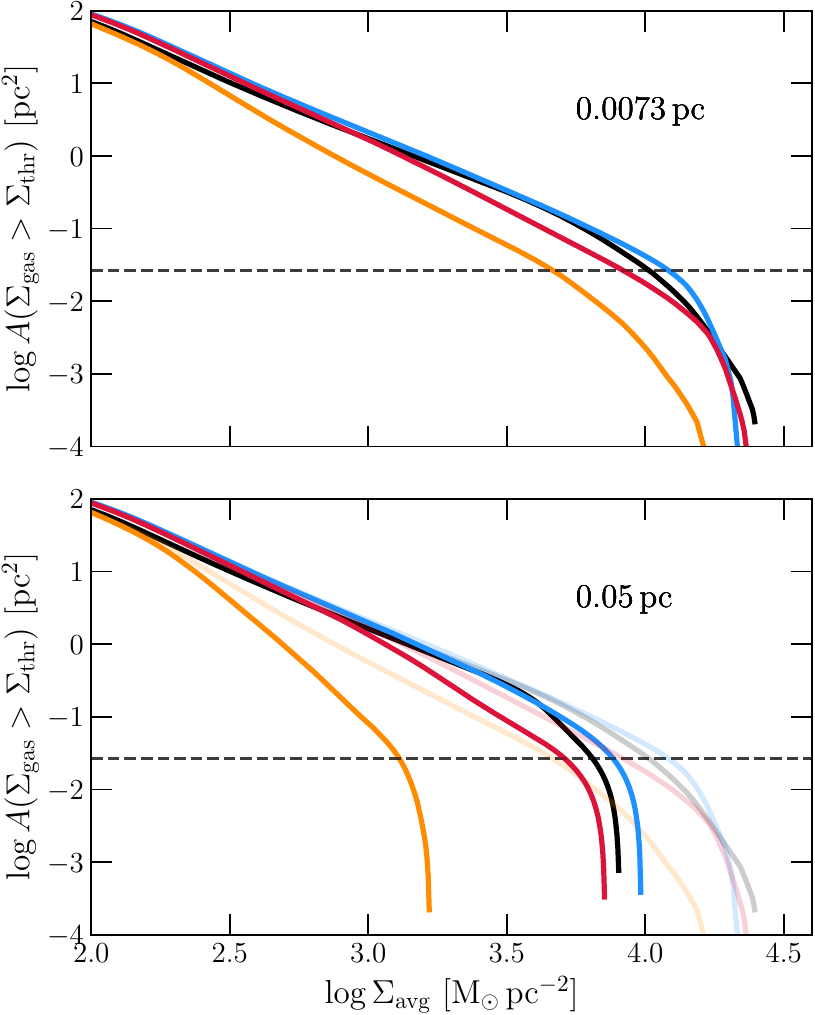}
    \caption{Cumulative area of the contours as a function of their average column density, at $3\,\Myr$. The two images are obtained from the original column density map (up) and the one smoothed with a Gaussian beam with $0.05\,\pc$ standard deviation (bottom). As in Figure~\,\ref{fig: biases}, the blended lines in the bottom panel are the ones plotted at $0.0073\,\pc$ resolution. The dashed line marks the area corresponding to the $500\,$pixels criterion.
    \label{fig: Cloud Structure}}
\end{figure}

To reproduce the second bias, we smoothed the column density maps, lowering the resolution of our ``observation''. This bias cannot impact the measured SFR, but rather affects the observed cloud structure\footnote{Resolution applied to YSOs detection constitutes instead a different issue. In high $\SigmaSFR$ zones, the crowding effect could lead to consistent decreases in completeness. \citet{2016AJ....151....5M, 2022PASP..134d2001M} showed how accounting for this the number of YSOs could increase by $30\%$ in the Orion complex.}.
Placing the cloud at a distance\footnote{That is the rough average distance of the clouds considered in \citep{2021ApJ...912L..19P} study.} of $650\,\pc$, we reproduced the physical resolution of the \textit{Herschel} satellite in the $500\,\mu\mathrm{m}$ band ($36\,$arcsec), convolving the column density maps with a Gaussian beam with a standard deviation of $0.05\,\pc$ (equivalent to a FWHM of $0.11\,\pc$). We did not modify the criterion given in Section~\ref{methods}, because now the threshold does not have to define a limit for the physical reliability of the data. Moreover, in this way, we could analyse how a low spatial resolution can affect the measured $\eff$ if this limitation is not carefully considered.

Smoothing the maps produces what is visible in the upper right graph of Figure~\ref{fig: biases}. Except \texttt{HIIR}, the curves of the simulations almost overlap with the original ones below $1000\,\coldensMPC$ (presented in Figure~\ref{fig: eff evolution}). Then, an increase is visible at the right end in \texttt{HIIR+PSJ} and \texttt{PSJ}, while this is evident at low $\Sigmaavg$ in the radiation-only run. This behaviour can be easily explained by examining how the function $A(\Sigmaavg)$ enters the relation. Combining Eq.~\ref{eq: dependences SigmaSFR},\,\ref{eq: modified KS} and \ref{eq: spherical approx}, we can express $\eff$ as\footnote{Notice that the usual $\tff\sim\Sigmaavg^{1/2}$ imply the presence of a constant scale width much smaller than the surface. While true when looking at external galaxies, this is no longer valid for clouds, where the third dimension is ``comparable'' to the others and its dependence must be included.} 
        \begin{equation}
            \eff = \SigmaSFR\frac{\tff}{\Sigmaavg}\sim\frac{\SFR}{A}\cdot \Sigmaavg^{-3/2}A^{1/4}\sim\SFR\Sigmaavg^{-3/2}A^{-3/4}.
        \label{eq: eff dependencies}
        \end{equation}
It follows that the cloud structure -- $A(\Sigmaavg)$ -- plays a central role in defining the behaviour of $\eff$ across the column density contours. As we approach the resolution limit, the integrated probability distribution function of  $\Sigmaavg$ (integrated N-PDF) drops sharply, leading to an increase in the efficiency per free-fall time thanks to the negative exponent of $A$. This causes the knee visible at high densities. 

In effect, by plotting the contour area as a function of their $\Sigmaavg$ at high and low resolution in Figure~\ref{fig: Cloud Structure}, we can identify the same feature. Moreover, we notice that the adopted threshold of $500\,$pixels (dashed line) effectively safeguards the data from resolution effects, as in the first graph the exponential decrease occurs beyond this point. Degrading the resolution moved this feature to lower $\Sigmaavg$, which caused the curves to bend toward the high-density end of Figure~\ref{fig: biases}.

However, this alone cannot explain the full picture.  Indeed, we see a steepening of the powerlaw tail of integrated N-PDF even at low densities. Technically, such variation should not be expected, as the smoothing should not significantly affect areas much larger than the beam size \citep[see\footnote{With little algebra, it is possible to demonstrate that the exponent of the N-PDF computed in logscale is the same of the one obtained from the integrated N-PDF \citep{2017A&A...606A.100L}.\label{ftnt: exponent NPDF}} e.g,][]{2010MNRAS.408.1089T, 2015A&A...575A..79S}. Most probably, the pitfall lies in the integration over the whole contour. As explained in Section~\ref{methods}, the method we applied does not take into account separated structures within the same column density level. Consequently, resolution can have a much stronger impact on the observed N-PDF of a highly fragmented structure. At $3,\mathrm{Myr}$, the \texttt{HIIR} cloud has been significantly restructured by the \hii region, resulting in the destruction of the central hub and the formation of several disconnected high-density clumps. The smoothing gives rise to a completely different $\gamma$ evolution.

\subsection{Instrument sensitivity}
\begin{figure*}
    \centering
     \subfigure[\label{fig: Lada compare a}]{\includegraphics[height=5.24cm]{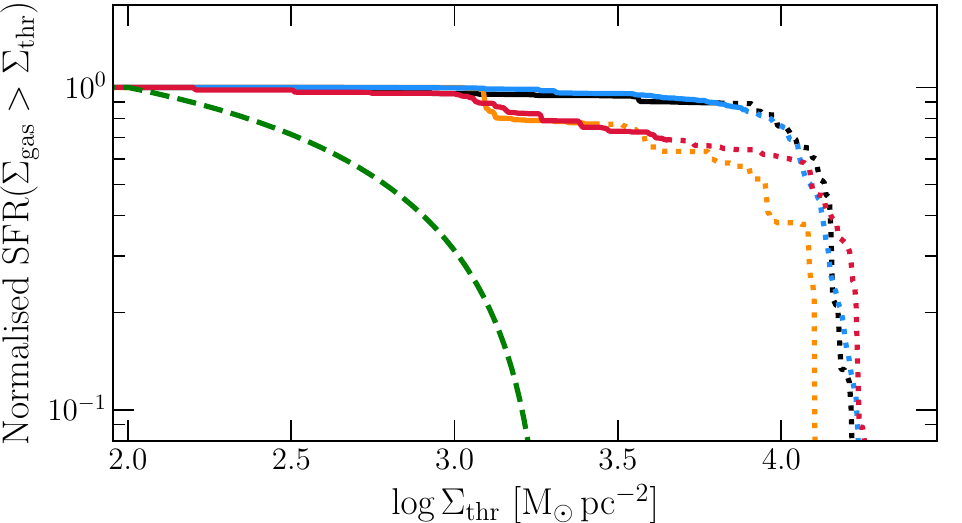}}
     \subfigure[\label{fig: Lada compare b}]{\includegraphics[height=5.24cm]{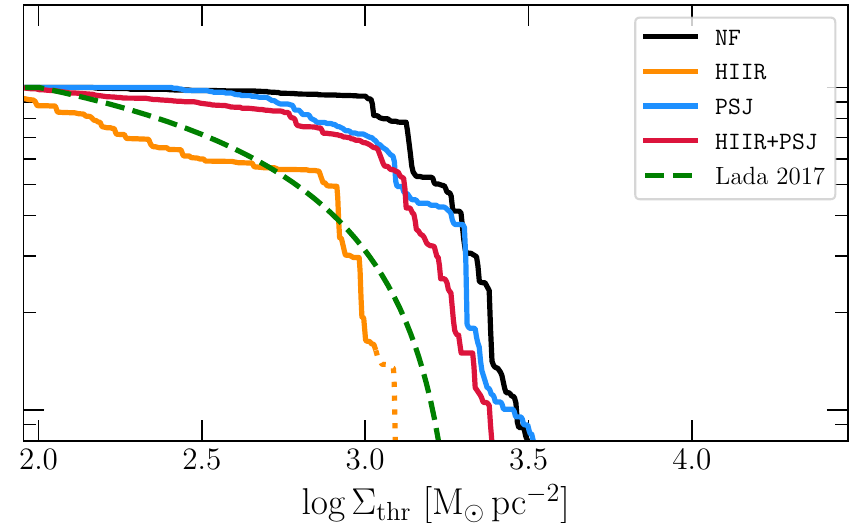}}
    \caption{SFR of density contours as a function of their threshold column density at $3.0\,\Myr$. The data are normalised to the total SFR at that time. The dotted part of the lines identifies isosurfaces with an area smaller than $500\,$pixels. The right graph shows the same distribution, but as it appears after the application of all three biases explained in the text -- YSOs counting for the SFR, resolution lowered to $0.05\,\pc$ and masking of extinct stars. The green dashed line reproduce the model from \citet{2017A&A...606A.100L}, with a threshold of $2000~\,\coldensMPC$ and a stellar probability distribution function of $\mathrm{PDF_*}\propto\Sigmathr^{-0.7}$.}
    \label{fig: Lada compare}
\end{figure*}
Finally, we accounted for the possibility that some YSOs may be too heavily extinct to be detected in the inner regions of the cloud. We considered the observations to be taken with the Spitzer IRAC $4.5\,\mu$m filter, which is extensively adopted in SFR measurements \citep[see e.g.][]{2009ApJS..181..321E, 2011ApJ...739...84G, 2013A&A...559A..90L}. 
For each sink, we evaluated its apparent magnitude as 
        \begin{equation}
            m_\mathrm{4.5}=M_\mathrm{bol}-\mathrm{BC}_{4.5}+5\log\left(\frac{d}{10\,\pc}\right) + A_{4.5},
        \end{equation}
where $M_\mathrm{bol}$ is the bolometric luminosity of the sink, $\mathrm{BC}_{4.5}$ its bolometric correction for IRAC $4.5\,\mu$m filter, $d$ the distance of the cloud, which we set to $650\,\pc$, and $A_{4.5}$ the extinction at $4.5\,\mu$m.

To recover the needed stellar properties such as the luminosity ($L$), radius ($R$) and bolometric correction, we compared the age and mass of each sink with the evolutionary tracks obtained from the MIST web tool\footnote{\url{ https://waps.cfa.harvard.edu/MIST/interp_tracks.html}}.
However, considering that in Class~0/I objects the luminosity due to gas accretion, $L_\mathrm{acc}$, is often a consistent fraction of their total luminosity $L_\mathrm{tot}$ \citep{2008A&A...479..503A, 2021A&A...650A..43F}, we corrected the sink's bolometric luminosity as $L_\mathrm{tot}=L+L_\mathrm{acc}$ \citep{2023ApJ...944..135F}. 
Following \citet{1998ApJ...495..385H}, we define $L_\mathrm{acc}$ as
            \begin{equation}
                L_\mathrm{acc} = 0.8\frac{GM}{R}\dot{M},
            \end{equation}
where $M$ and $\dot{M}$ are the sink mass and its accretion rate averaged over $\approx0.05\,\Myr$. 
Finally, we obtained the last ingredient $A_{4.5}$ evaluating the amount of gaseous material in front of the sinks, with the conversion $A_{4.5}=0.572\cdot A_\mathrm{K}$ provided in \citet{2013A&A...549A.135A}, and adopting $\Sigmagas=197\,\coldensMPC\cdot A_\mathrm{K}$ \citep[hence, $\Sigmagas=384\,\coldensMPC\cdot A_{4.5}$;][]{2013ApJ...778..133L}.

To be detected, we required sinks to have $m_{4.5} < 14.5\,$mag, which is the sensitivity limit at $4.5\,\mu$m reported in the Spitzer/IRAC Candidate YSO catalogue \citep[SPICY catalogue;][]{2021ApJS..254...33K}. However, we stress once again that, with these simplified prescriptions, our goal is to understand how those biases could affect the observed $\eff$--$\Sigmaavg$ relation. Indeed, observers employ different instruments at different wavelengths to limit the impact of these biases \citep{2007ApJ...663.1149H, 2016AJ....151....5M}. These corrections are particularly demanding to reproduce in simulations and require specific treatments that are beyond the scope of our study.

The effect of the instrument sensitivity is shown in the third graph of Figure~\ref{fig: biases} (lower left). A notable feature is the reduction of $\eff$ average value.
Even if caution has to be used in the interpretation, the resolution of the tension with \citet{2021ApJ...912L..19P} data could imply that the physics implemented in \texttt{PSJ} and \texttt{HIIR+PSJ} is sufficient to describe the mechanisms involved in setting the $\eff$ in molecular clouds. In our simulations, adding a detection threshold has a strong impact on the observed SFR, leading to an underestimation of a factor of 3 of $\eff$ in the full-feedback run, and $\approx5$ in the jet-only run. This difference is solely caused by the different distribution of YSOs, as supported by the fact that this bias has a softer impact on the \texttt{HIIR} simulation. Indeed, by $3\,\Myr$ the main \hii had already significantly restructured the cloud, dispersing a consistent part of the gas, reducing the extinction experienced by stars. 

Another consequence of this bias is the rapid fall of the relation at high $\Sigmaavg$. Being a log-log graph, the slope of $\eff$ is directly affected by the fractional variation of $\SFR$ at each contour. Most of the stars -- and so the majority of the SFR -- reside at high column densities. Therefore, all low-$\Sigma$ contours have comparable $\SFR$. Similar drops are also observed in \citet{2021ApJ...912L..19P} clouds when $\Sigmathr$ exceeds their reliability criterion (defined by the column density at which $\tff>0.5\,\Myr$). It is important to highlight that this reduction is specific to the \textit{cumulative} method. Since $\SigmaSFR$ is obtained through the integration over the entire contour, missing YSO in the central regions lowers the SFR of every contour, even those with low $\Sigmaavg$. In contrast, a \textit{differential} approach like that used in \citet{2014ApJ...782..114E} would not be affected, but it would require rougher assumptions for computing $\tff$ and would suffer from the lower statistics of YSOs found at low densities.

We can show this phenomenon more accurately if we extract the relation between $\SFR$ and the contour column density. In Figure~\ref{fig: Lada compare a}, we plotted the evolution of $\SFR$ for different contour levels at the maximum resolution, as a function of the column density threshold of that contour $\Sigmathr$. With this change of $x$-axis, we can provide a comparison with the model by \citet{2013ApJ...778..133L}, shown in the graph as a green dashed line. According to this work, the SFR included in each contour can be described as:
        \begin{equation}
            \SFR=\alpha_\mathrm{N}\int_{\Sigmathr}^{\Sigmamax}  \mathrm{PDF}_*\,\der\Sigmagas\,,
            \label{eq: integrated SFR}
        \end{equation}
where $\alpha_\mathrm{N}$ is the normalisation constant and $\mathrm{PDF}_*$ is the probability of finding a star at a given gas column density. As derived from the California Cloud in their work, we use 
        \begin{equation}
                \mathrm{PDF}_\star \propto \begin{cases}
                \Sigmagas^{\phi} &\text{if $\Sigmagas>\Sigmamin$}\\
                0 &\text{otherwise}
                \end{cases}
                \label{eq: lada}
        \end{equation}
with $\phi=-0.7$ and $\Sigmamin=100\,\coldensMPC$, and integrate up to a column density $\Sigmamax=2000\,\coldensMPC$ ($A_\mathrm{K}\approx 10\,$mag).  

The net differences between our simulations and Eq.~\ref{eq: lada} are clear and quantitative. In fact, attempting a fit leaving $\Sigmamax$ and $\phi$ as free parameters and $\Sigmamin$ equal to the threshold at which the first accreting star is removed, leads to $\phi\approx1$ in all the simulations, while for the four clouds studied \citet{2017A&A...606A.100L} this varies between -1 and -0.3.
This picture changed after the application of all the observational biases described above (Figure~\ref{fig: Lada compare b}). 
Qualitatively, the curves now resemble much more closely the model. Repeating the fit of Eq.~\ref{eq: lada}, we found that only \texttt{HIIR+PSJ} and \texttt{PSJ} converged to reasonable values of $\phi$ ($-0.32\pm0.02$ and $-0.42\pm0.02$ respectively)\footnote{Run \texttt{NF}, instead, returns a value of $\phi=0.68\pm0.03$, while in run \texttt{HIIR} the fit does not converge.}.

We can evaluate the derivative of Eq.~\ref{eq: integrated SFR} to understand how the sensitivity changes the contribution of $\SFR(\Sigmathr)\sim\SFR(\Sigmaavg)$ to the observed KS. Assuming that Eq.~\ref{eq: lada} holds at every density contour, with little algebra we arrive at
        \begin{equation}
            \frac{\der \ln \SFR}{\der \ln \Sigmathr}(\Sigmathr) = -(\phi+1)\frac{\chi^{\phi+1}}{1-\chi^{\phi+1}}
        \end{equation}
with $\chi=\Sigmathr/\Sigmamax$. At lower densities, two ingredients contribute to the initial slope of $\SFR$. The first is exponent $\phi$, that \citet{2013A&A...559A..90L} showed to be the sum between the exponent $-q$ of the $A(\Sigmaavg)$ function and the exponent of the KS relation as obtained from the differential approach. The steeper this last relation, the more stars will be present at higher densities and hence the shallower will be the logarithmic derivative. The second is $\chi$, which is the combination of the scale at which Eq.~\ref{eq: lada} ceases to be valid, and the sensitivity limit of the instrument.

\subsection{Biases combined}
To conclude, in the last panel of Figure~\ref{fig: biases}, we plotted $\eff$ as seen after the application of all these biases on our column density maps. First, we masked sinks too extinct to be detected. Then, we smoothed the column density maps. Finally, we repeated the same analysis described in Section~\ref{methods}, using the new YSO count prescription to evaluate $\SigmaSFR$.

The resulting relations are similar to what we would obtain from the simple sum of all the effects, meaning that any coupling between the biases has little effect on the resulting relation. However, a notable exception arises when looking at the curve of \texttt{NF}. In this panel, $\eff$ is much higher than the one observed in the ``Sensitivity'' graph. This cannot be due to the smoothing of the column density map, as we said that this does not affect the total SFR detected. Therefore, the YSOs counting must increase the observed SFR, although the first panel clearly shows that this decreases the average $\eff$ when taken alone. 

The answer resides in the same argument we used in Section~\ref{subsec: yso counts}. Counting YSOs accurately reproduces the SFR only when the correct average mass is used. However, if the mass distribution varies as we approach the densest regions of the cloud, then a new bias is introduced. In effect, as we increase $\Sigmathr$ we observe a consistent increase in the average stellar mass of the contours.
In \texttt{NF}, the average sink mass in the hub increases significantly, thus $\langle m_\star \rangle$ overestimates the mean mass of YSOs found at low densities while underestimating the ones found at the highest ones. Hence, if we remove the stars in the centre of the cloud, this approximation tends to overestimate the SFR observed.

We remark that in our cloud, this phenomenon is probably attributable to numerical effects (see Section~\ref{methods}), since in all the simulations containing feedback $\langle m_\star \rangle$ remains fairly constant even at high $\Sigmathr$. Nevertheless, this offers insights into the interpretation of observations as well. Whether the mass function varies as a function of the local density is a subject still under debate \citep{1998A&A...333..897R, 2002A&A...394..459S, 2008ASPC..390...26B}. Moreover, in young massive clusters, the mass segregation process takes place in very short timescales, possibly leading to an effect of this kind \citep{2010ARA&A..48..431P, 2010MNRAS.407.1098A, 2022A&A...667A..69S}.

\section{On the shape of the relation} \label{dis}
\subsection{Cloud structure}
\label{sec: cloud structure}

\begin{figure}
    \centering
     
     \includegraphics[width=1\linewidth]{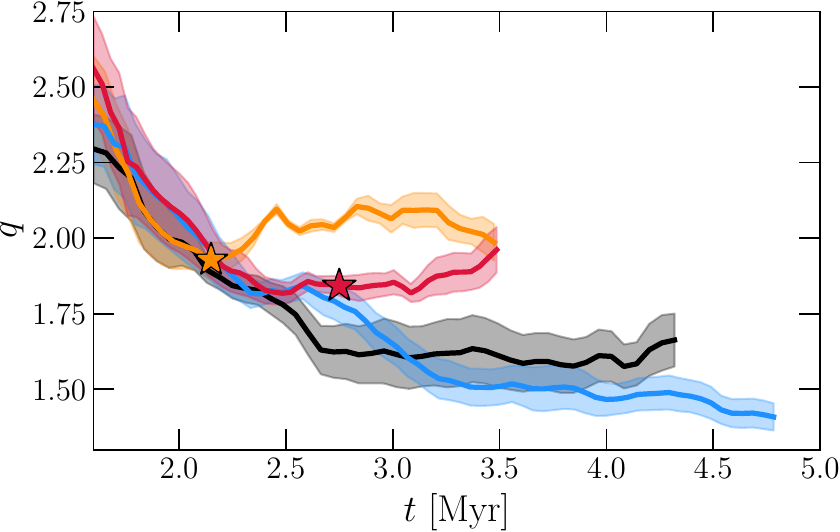}
    \caption{Temporal evolution of the fitted exponent $q$, averaged over the three axes, in the four simulations. The colour code is the same as in Figure \ref{fig: SFR glob}. The shaded areas represent the standard deviation of the sample due to the differences between the lines of sight. The uncertainties on the single fits are negligible. As in Figure~\ref{fig: SFR glob}, the stars highlight the moment of the eruption of the main \hii region. 
    \label{fig: fit CloudS}}
\end{figure}
In the previous Section, we have found clues that the function $A(\Sigmaavg)$ plays a central role in shaping the relation. Figure~\ref{fig: Cloud Structure} demonstrated that the effects produced by low resolution on the observed cloud structure directly translate into similar features in the $\eff$ versus $\Sigmaavg$ graph. Simultaneously, Figure~\ref{fig: Lada compare} has shown that the more we can access the SFR in the highest densities of the cloud, the more $\SFR$ appears flat at lower densities. This is because most of the SFR happens in the densest regions.
Consequently, the shape of the relation at low-mid densities is mostly governed by the slope of the $A(\Sigmaavg)$, while at high densities it intertwines with the loss of stars between contours. Which of the two dominates in this regime may vary from cloud to cloud (e.g.~in the last panel of Figure~\ref{fig: biases}, compare \texttt{HIIR} with the other runs).

In Section~\ref{subsec: resolution}, we already found that the presence of ionising feedback does play a role in restructuring the cloud, changing the observed area-to-column-density relation. Even at the maximum resolution, in Figure~\ref{fig: Cloud Structure}, some differences are evident between the various setups. In the radiation runs, the compression caused by the expansion of the \hii region fronts is not sufficient to overcome the gas dispersion, and comprehensively the total amount of dense gas decreases, leading to a steeper relation. To measure this effect quantitatively, we fitted a powerlaw relation
        \begin{equation}
            A\sim \Sigmaavg^{-q},
            \label{eq: cloud structure}
        \end{equation}
at each snapshot. We performed the fit with the \texttt{Python} built-in function \texttt{curve\_fit}, using only the contours having $\Sigmaavg>400\,\coldensMPC$ ($\log\Sigmaavg\approx2.6$) and an area greater than 500 pixels. The first constraint allowed us to exclude from the fit the largest scales of the simulations, which are dominated by the global collapse and thus display an exponent $q$ similar to that of \texttt{NF}.
In Figure~\ref{fig: fit CloudS}, we plotted the fit results, averaged over the three axes of projection. The uncertainties associated with the fit are negligible, so the shaded area represents the dispersion between the three lines of sight. Without any support from protostellar jets, the runs \texttt{NF} and \texttt{HIIR} collapsed faster and achieved lower values of $q$ (more dense gas at high $\Sigmaavg$). The onset of large \hii regions froze the exponent (at $\approx2.3\,\Myr$ in \texttt{HIIR} and $\approx2.8\,\Myr$ in \texttt{HIIR+PSJ}), which settled to a value of roughly 2 (>1.85 in \texttt{HIIR+PSJ}, $\approx2.05$ in \texttt{HIIR}). 

Conversely, in \texttt{NF} the exponent continued to decrease down to roughly 1.6, close to what is reported and expected in gravoturbulent simulations without feedback \citep{2015ApJ...808...48B, 2017ApJ...834L...1B, 2018ApJ...859..162C}. Notably, the run with only jets achieves a shallower relation than the run with only gravity, reaching values as low as 1.4. We notice that the values of $q$ between \texttt{NF} and \texttt{PSJ} cross almost exactly at $3\,\Myr$, so that the steeper relation is not visible in Figure~\ref{fig: Cloud Structure}. However, looking at Figure~\ref{fig: SFR glob}, we see that this excess of dense gas is not rewarded by an increase in the total SFR. This is probably because compressed material is ejected at high relative velocities relative to the surrounding gas flow. 

We can compare these findings with what is usually observed in real molecular clouds \citep[cf.\footnote{Notice that \citet{2021A&A...653A..63S} fits the exponents $\alpha$ of the N-PDF in logspace. Hence, their exponent $\alpha$ is identical to our $q$ (Footnote\,\ref{ftnt: exponent NPDF})}][]{2015A&A...576L...1L, 2021A&A...653A..63S}. Our simulated clouds attained slopes similar to the observed ones, although it appears that values of $q$ near or above 2 are more frequent in our Galaxy. 
Moreover, the fact that only clouds with radiation attained a similar exponent may represent a central result of this work. Indeed, \citet{2015A&A...581A..74A} conducted an extensive study on the N-PDF distribution in 195 Galactic molecular clouds, concluding that clouds with \hii regions naturally develop a powerlaw tail with an index -2. The authors suggested this slope to be caused by the formation of an isothermal-sphere-like structure. Indeed, based on the work from \citet{2011ApJ...727L..20K}, they showed that for a spherical density distribution of the form $\rho_\mathrm{gas}\sim r^n$, with $r$ the distance from the centre, a relation between $n$ and $q$ can be established, such that $n = 1 + 2/q$. Setting $q=2$, we recovered the isothermal sphere distribution. Looking at Figure~\ref{fig: Cloud Structure}, the presence of a mechanism that pushes $q$ towards this value is also suggested by the smaller dispersion between the line of sights that characterises the radiative runs (although a larger sample of simulation would be needed to confirm this result statistically). This could potentially mean that radiation feedback plays a fundamental role in setting the density structure of clouds. This could provide valuable insights to distinguish between different classes of cloud.

Interestingly, the same study from \citet{2015A&A...581A..74A} detected that on average the star-forming clouds observed without \hii regions display slopes steeper than the one reported here. This seems to contrast with our results. Nonetheless, such a steep slope is visible at the early stages of the simulations (left-hand side of Figure~\ref{fig: Cloud Structure}). Thus, a possible explanation is that the clouds that have such a high $q$ are simply younger than the others. Possibly, those clouds will proceed to evolve a shallower slope, but eventually form a massive star that interrupts the process as happened in \texttt{HIIR+PSJ}.

\subsection{Impact of feedback}
Figure~\ref{fig: fit CloudS} gives us the necessary instruments to interpret the behaviour of the observed KS relations in Figures~\ref{fig: SFR evolution}-\ref{fig: eff evolution}, since we found in the previous Section that $\SFR$ has very little impact at low column densities. In the two Figures, the radiative simulations show a slope nicely falling onto a $\SigmaSFR\sim\Sigmaavg^2$ in the first and $\eff\sim const$ in the second. This behaviour can be perfectly explained by the reported $q\approx2$. Concerning the KS relation, combining Eq.~\ref{eq: dependences SigmaSFR} with the $A(\Sigmaavg)$ in Eq.~\ref{eq: cloud structure}, it is clear that a nearly constant $\SFR$ leads to $\SigmaSFR\sim\Sigmaavg^q$. Doing the same for Eq.~\ref{eq: eff dependencies}, it is easy to find that under the same $\SFR\sim const$ approximation, a value of $q\approx2$ leads to a constant
$\eff$. Looking closely we can find that Figures~\ref{fig: eff slope}-\ref{fig: fit CloudS} closely resemble each other. At the same time, the weaker dependence on $A$ of Eq.~\ref{eq: eff dependencies} allows the relation to be more resilient to the effect of feedback, explaining the smaller deviations between the simulations compared to the differences found in the KS relations (see Section~\ref{sec: cloud structure}).

This also suits the shallow behaviour of \texttt{HIIR} and \texttt{HIIR+PSJ} in the 3D analysis (Appendix~\ref{app: 3d-2d}; Figure~\ref{fig: SFR 3d}). With little algebra, it is possible to show that if the same assumption of spherical symmetry is valid, then $V\sim\rho_\mathrm{gas}^{-\frac{3q}{2+q}}$. If $q\approx2$, the influence of the volume in the tri-dimensional version of Eq.~\ref{eq: eff dependencies} is expected to be negligible, since
        \begin{equation}
            \epsilon_{\mathrm{ff},\,3\mathrm{D}}=\frac{\SFR}{V\rho_\mathrm{gas}}\tff\sim \SFR\rho^{\frac{3q}{2+q}-\frac{3}{2}},
        \end{equation}
with the exponent going to zero for $q=2$.

Summing up all these findings, we conclude that when ionising radiation is included in the simulation, the cloud systematically develops a KS relation more weighted toward higher column densities, as well as a higher $\eff$ and $\epsilon_{\mathrm{ff},\,3\mathrm{D}}$ -- although with a reduced impact. Thus, although radiation negatively impacts the SFR on the largest scale and reduces the amount of dense gas available, the remaining dense gas manifests a higher $\eff$.

Unfortunately, the interpretation of the observed trend is not unique. One possible explanation is that the expansion of ionised gas promotes star formation at high densities in an active way. In this scenario, the radiation front would create favourable conditions for the development of new star formation sites, either compressing the surrounding gas until it becomes gravitationally unstable \citep[collect and collapse mechanism;][]{1977ApJ...214..725E}, creating converging shocks at clump surfaces capable of enhancing significantly their core density \citep[radiation driven implosion;][]{1982ApJ...260..183S}, or acting as a compressive driver for local turbulence, \citep[as suggested by][]{ 2020MNRAS.493.4643M, 2021MNRAS.500.1721M}.
Alternatively, it could be suggested that massive stars wipe out only the gas that is not sufficiently gravitationally bound, leaving only the dense gas that would have formed stars regardless. This again would lead to deriving a higher SF efficiency in the region, since the local SFR would roughly remain the same, but the observed amount of dense gas would be lower.
Which of the two processes prevails is unclear and subject to intense studies in the field. Either way, the outcome of this interplay produces observable features in our cloud, which could aid in interpreting observational data. For instance, in a sample of real clouds, the differences in their evolutionary stage or variations in the feedback mechanisms' strengths could contribute to the observed scatter in the derived KS laws.

Indeed, Figures~\ref{fig: SFR evolution}-\ref{fig: eff evolution} show that the relations undergo significant variations over a timescale of a few million years, consistently with what is reported in \citet{2018MNRAS.475.3511G}. As time progresses, the four extracted relations intersect at different column densities. This has implications for observational studies that extract single measurements of $\eff$ or $\SigmaSFR$ from each cloud. These analyses often adopt a common minimum $\Sigmathr$ to define the extent of the clouds. Our results emphasize that achieving similar values of $\eff$ or $\SigmaSFR$ at a specific column density contour does not necessarily indicate similarities in the underlying physical conditions.

To further illustrate this point, we mimicked the analysis on an observational set of star-forming complexes, treating each snapshot as an observation of a distinct cloud. We did not apply the biases of Section~\ref{sec: biases}, to evince conclusions that do not depend on the assumptions made in that Section. For each snapshot, we extracted the value of $\eff$ corresponding to a given column density contour. We chose five different column densities as reference points: at $200,\,500,\,1000,\,2000$ and $5000\,\coldensMPC$. Consequently, we obtained four datasets (one for each feedback prescription) and computed the temporal average and dispersion of $\eff$ for each simulation. The question is whether we can distinguish the different physical conditions of the clouds when we lack information about their temporal evolution and the dependence on $\Sigmaavg$.

\begin{figure}
    \centering
     
     \includegraphics[width=1\linewidth]{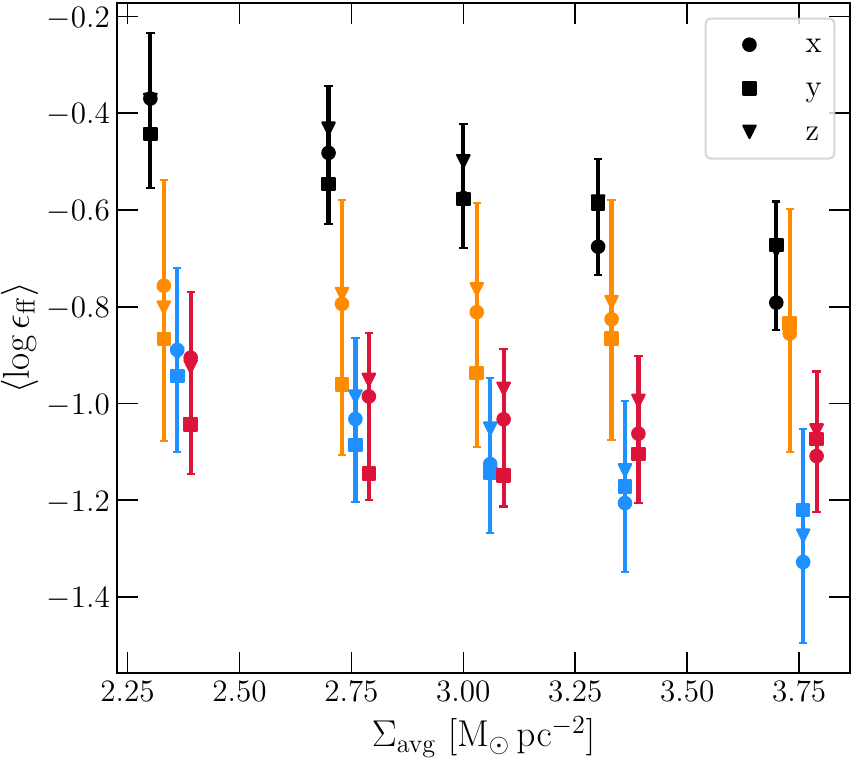}
    \caption{Time average of $\log\eff$ at 250, 500, 1000, 2500, 5000$\,\coldensMPC$ in the various simulations. The data has been slightly displaced to make the comparison clearer. The different markers highlight different axes of projections, while the error bars show the dispersion of $\eff$ measured at a given column density, due to its evolution across the simulation. The effects from the line of sight and stellar feedback are not enough to overcome this dispersion, so that in a set of clouds at different evolutionary stages those will not be visible if the dependence from $\Sigmaavg$ is not taken into account.}
    \label{fig: eff avg}
\end{figure}

The results, presented in Figure~\ref{fig: eff avg}, reveal that, except for \texttt{NF}, the averages of the different simulations are indistinguishable because of their significant temporal variations. If these were actual observations, the impact of stellar feedback would always be secondary compared to the variation due to the cloud evolution. This does not mean that feedback has no valuable impact on the SFR, since, looking at the full picture (Figure~\ref{fig: eff evolution}), $\eff$ changes substantially with $\Sigmaavg$ when the feedback prescription is changed. Therefore, the cause is that $\eff$ is highly varying over time and that the different runs achieve the same $\eff$ during their evolution, but at different times.  Only when looking at $\eff(\Sigmaavg)$ a clear dependence on the feedback mechanism becomes evident.

In addition, we notice that the differences just mentioned are mostly visible above the range of densities typically studied. Studies such as \citet{2015A&A...576L...1L}, \citet{2021ApJ...912L..19P} or \citet{2022ApJ...938..145B} reach maximum column densities of the order of $1000\,\Sigmaavg$, while the high resolution of these simulations allows us to probe an order of magnitude more. This density range is mainly governed by the cloud's structure, and observations may not be able to catch variations caused by feedback mechanisms. 

Therefore, what is most interesting for observational datasets we can get with the present instrumentation are the differences in the slopes, which are detectable even at lower densities. From our study, we would expect clouds with strong radiation feedback to display the steeper slopes in the $\SigmaSFR$ versus $\Sigmaavg$ plot. It is extremely interesting to notice that in \citet{2021ApJ...912L..19P}, clouds dominated by strong radiative feedback (Orion A, Sh2-140, Cep OB3, and Cygnus X) show on average a steeper slope than the other ones (Ophiucus, Orion B, Perseus, Aquila North, Aquila South, NGC 2264, AFGL 490, and Mon R2). In particular, the mean exponent of the KS relation found for the first clouds is $2.24$ with a dispersion of $0.14$, while the average for the second group is $1.87$, with a dispersion of $0.24$. Although the sample is limited, this supports our findings, and further studies that extend the sample could provide full confirmation of the trend.

\section{Conclusions} \label{conc}
We conducted an in-depth analysis of high-resolution MHD simulations to investigate the impact of stellar feedback on two different star formation models: the original KS law and the \citet{2005ApJ...630..250K} $\eff$ model. Applying techniques commonly used in observational works, we revealed that substantial differences arise when stellar radiation and protostellar jets act to reshape the cloud structure. In particular, we found that the coupling between area-column density relation and feedback mechanisms plays a major role in the observed star formation law, much more than the stellar distribution itself. 

When only protostellar jets are active, they compress part of the gas so that the N-PDF becomes more weighted toward high densities. However, this does not translate directly into an increase in the SFR, resulting in a shallower KS law and a lower $\eff$ at higher column densities. On the other hand, \hii regions act dispersing the gas, leading to a steeper $A$ versus $\Sigmaavg$ relation, with an exponent that settles around a value of $q=2$. This particular value also validates the study from \citet{2015A&A...581A..74A} conducted on Galactic molecular clouds. Even though the presence of ionising stars decreases the global SFR, the steepness of the $A(\Sigmaavg)$ function is such that the observed star formation efficiency is higher at high column densities. The result is confirmed even for $\eff$, although the weaker dependence on $A$ mitigates the differences with the only jet run.

We also applied prescription to emulate the effect of three different biases observations have to deal with. While assigning the same mass and age to each YSO detected does not significantly alter the observed SFR, the presence of a detection threshold to embedded YSOs and the limited resolution of column density maps can modify the shape of the relation at the highest densities. In particular, we showed that low-resolution column density maps can affect the $A$--$\Sigmaavg$ relation when observing a highly structured cloud. In addition, we report a significant impact of the presence of a detection threshold, which both leads to a steep drop in the observed $\eff(\Sigmaavg)$ relation at high densities and to an underestimation of the global SFR. 

When applied altogether, the effect did not couple efficiently so the resulting relation was close to the one obtained by simply summing the biases. The only exception was \texttt{NF}, where the mass distribution of sinks changed toward the centre of the cloud. This caused an overestimation of the SFR due to the intertwined action of the YSOs counting and the lower completeness of the detected embedded sources. Although this could be attributed to numerical effects, it could play a role while observing regions where the mass segregation mechanism is ongoing (either dynamical or due to the initial cluster formation). When all the biases are applied, we recover a good agreement with the observational results from \citet{2021ApJ...912L..19P}, which suggests that the physics included in our simulation is enough to describe the process of star formation at the cloud scales.

Finally, having access to the temporal evolution of our cloud, we could show that the exponent of the KS relation and the value of $\eff$ can vary significantly with time. This means that in a sample of real clouds, where different evolutionary stages are present simultaneously, extracting the value of $\eff$ at a single column density would hide the effect of feedback. Indeed, distinct departures between the models become only evident when considering the dependence on $\Sigmaavg$. 

We will extend the sample of numerical simulations to cover a wider range of initial conditions and feedback strengths, to parameterise the impact of ionising stars as a function of cloud and stellar properties, and to perform a direct comparison with observational data. 

\vspace{0.5cm}
      We thank Mark R.~Krumholz for the valuable insights provided. A.Z. thanks the support of the Institut Universitaire de France (IUF). We thank the referee for his relevant and focused insights, which have helped us enhancing the contextual relevance and clarity of the developments outlined in this work.
      This research has received funding from the European Research Council synergy grant ECOGAL (Grant : 855130).

  
\bibliographystyle{aa} 
\bibliography{SFR} 
\appendix
\section{Effect of projection axis}
\begin{figure*}[h!]
    \includegraphics[width=0.99\linewidth]{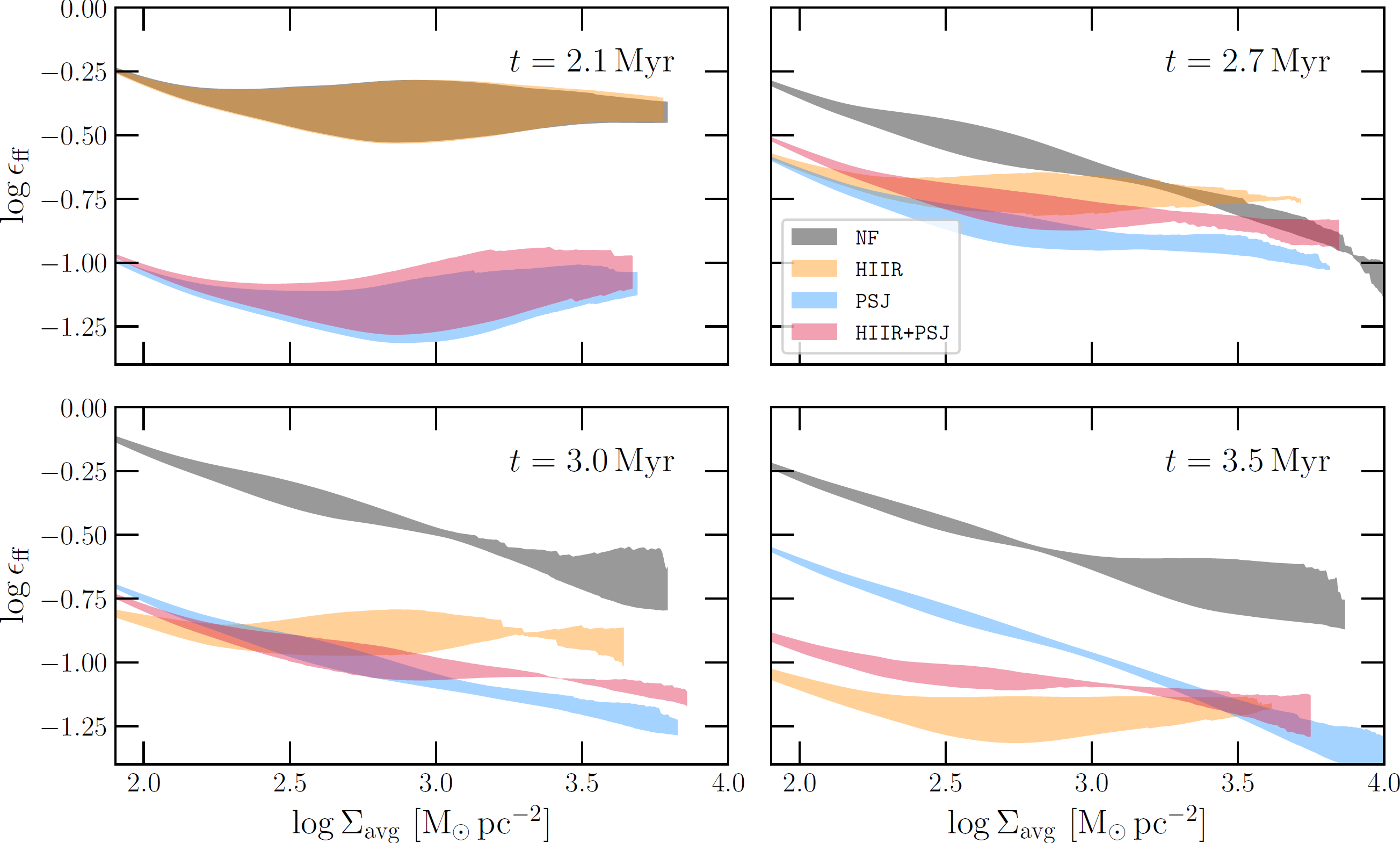}
    \caption{Same as Figure~\ref{fig: eff evolution}, but using the average and $1\,\sigma$ dispersion of $\eff$ obtained from the three lines of sight. Data are masked when at least one of the three contours at the same density threshold does not satisfy the 500\,pixels criterion. 
    \label{fig: eff axes}}
\end{figure*}
In our analysis we mainly showed the results from the projection along the $z$-axis. However, the other axes behave similarly and no notable difference has to be reported. Figure~\ref{fig: eff axes} shows how the KS relation changes when averaging between the three lines of sight. The shaded areas represent the $1\,\sigma$ dispersion of the three curves. It is visible that the variability decreases with time, and the results remain coherent. This could also be guessed looking at Figure~\ref{fig: eff slope}, where the dispersion of the fitted slope is small.
\label{app: los}
\section{2D versus 3D star formation relation}
\begin{figure*}[b!]
    \centering
    {\includegraphics[width=0.521\linewidth]{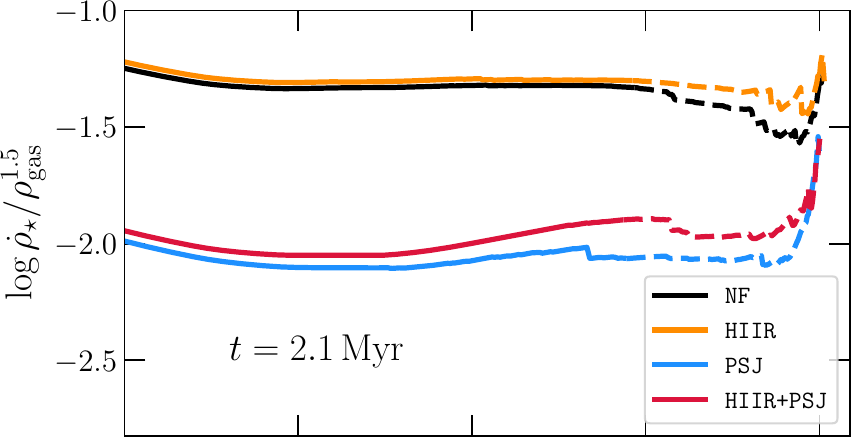} 
    }
    \hfill\raisebox{-0.0295cm}{
    {\includegraphics[width=0.45\linewidth]{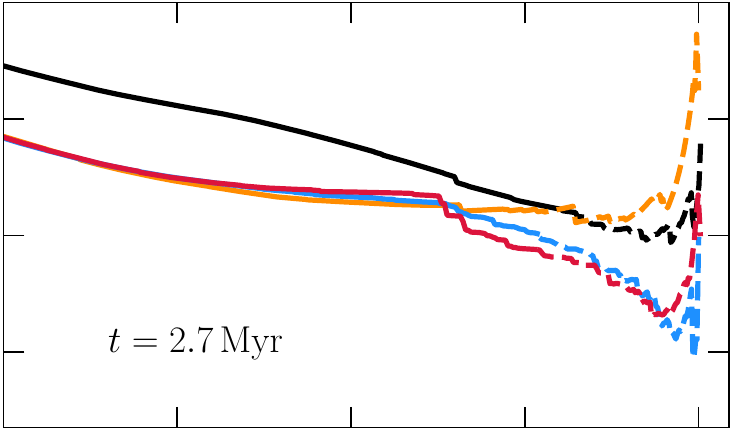}}
    }
    \\\vspace{0.36cm}
    {\includegraphics[width=0.52\linewidth]{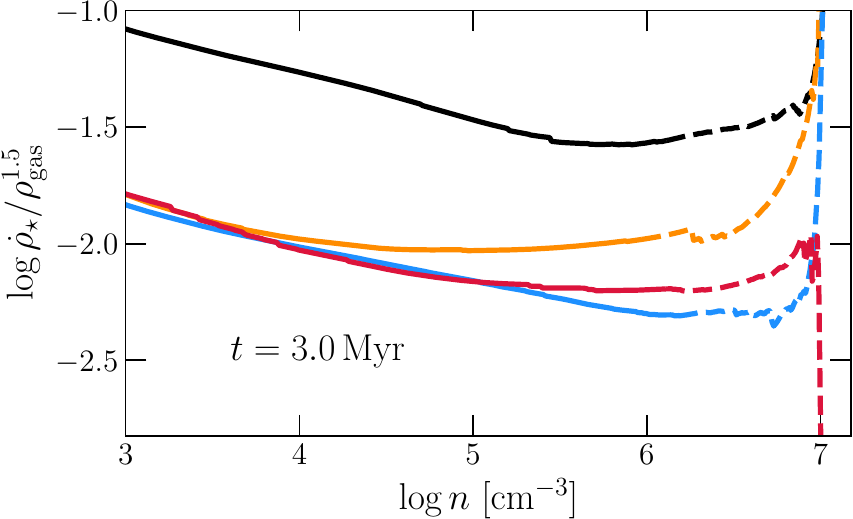}   
    }
    \hfill\raisebox{-0.0cm}{
    {\includegraphics[width=0.45\linewidth]{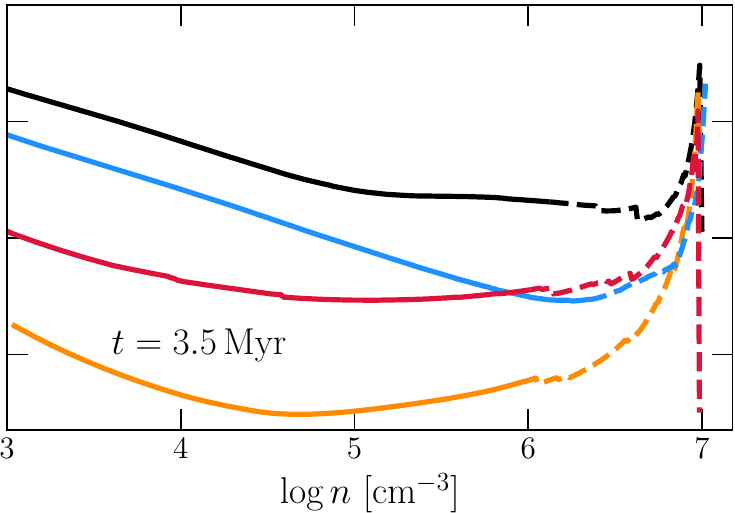}}
    }   
    \caption{Evolution of  $\dot{\rho}_\star$ as a function of volume number density. The dashed lines mark the 3D contours with volumes smaller than $4\cdot10^{-3}\pc^3$. Units of the $y$-axes are $\MSun^{-0.5}\pc^{1.5}\Myr^{-1}$. The steep rise in the right-hand side is caused by the density threshold for sink formation at $n=10^7\,\ndens$.
    \label{fig: SFR 3d}}
\end{figure*}

The non-trivial correlation between local quantities and their projected value constitutes a major challenge to the interpretation of astrophysical observations.
With simulations, one can accurately preserve the evolution of physical variables for each cell, accessing the full 3D information. Therefore, it is of great interest to understand the extent to which projection effects impact the results. Hence, we extended the same cumulative method described in Section~\ref{methods} to perform the three-dimensional analysis. We divided the cloud into 500 volume density contours, evenly spaced in log between $10^3$ and $10^7\,\ndens$. Converting the 500\,pixels criterion used for 2D contours, we require the isosurfaces to contain at least a volume greater than $500^{3/2}\cdot (7.3\cdot10^{-3}\pc)^3\approx4\cdot10^{-3}\pc^{3}$. We plotted the outcome in Figure~\ref{fig: SFR 3d}, where now we scaled the the $y$-axis (the volume SFR density $\dot{\rho}_\star$) by $\rho_\mathrm{gas}^{1.5}$, where $\rho_\mathrm{gas}$ is the average volume density of the contour. The choice of the exponent is justified by Figure~\ref{fig: eff evolution}. Since Eq.~\ref{eq: modified KS} display a constant behaviour with $\Sigmaavg$ in the radiative simulations, we expect a similar feature when dividing by $\rho_\mathrm{gas}^{1.5}\propto\rho_\mathrm{gas}/\tff$.

The first thing we could estimate from the plots is the limit up to which data can be considered reliable. Indeed, we expect the efficiency to increase sharply as we approach the density threshold for sink formation \citep{2021MNRAS.507.4335K}, and we can see a steep the rise  towards the very high-density edge of the graph. Keeping in mind that caution is always necessary when comparing three- and two-dimensional datasets, the distance between the knee of the rise and our reliability limit (beginning of the dashed line) allowed us to be confident that our constraint provides sufficient protection against resolution effects. 

Comparing these graphs with their respective counterparts in Figure~\ref{fig: SFR evolution}, we can confirm that the features observed in the projection are supported by the 3D analysis. The plot at $2.1\,\Myr$ shows that its 3D version closely resembles the two-dimensional version. At $t=2.7\,\Myr$, runs with jets exhibit similar trends. At the same time, the radiation-only simulation presents a comparable $\dot{\rho}_\star$ to the run with both jets and radiation at low densities, but an higher efficiency at higher densities. Graphs at $3.0$ and $3.5\,\Myr$, instead, require more attention. Indeed, the snapshot at $3.0\,\Myr$ confirms the findings of the previous section, with the relative behaviours that remain preserved, even if the differences appear to be smaller. In particular, one may notice that the departure of \texttt{HIIR+PSJ} from \texttt{PSJ} happens ``later''. However, comparing points between the two graphs is dangerous because, without a characteristic scale, there is no exact conversion from column to volume density. 

At $3.5,\Myr$, the slope displayed by simulations with radiation flattened out even more. However, there are some notable differences. In the \texttt{HIIR} run, the star formation event has nearly ceased, and the SFR density curve lies below the others. On the other hand, the \texttt{HIIR+PSJ} run achieves a $\dot{\rho}_\star$ comparable to \texttt{PSJ} at high column densities $n$, although this happens just below the reliability limit. Nevertheless, this does not affect the conclusions drawn so far. Indeed, at other times in the simulations, the observed trends are consistent with the previous discussions.

\label{app: 3d-2d}

\end{document}